\RequirePackage[hyphens]{url}

\newif\ifinappendix

\documentclass[compsoc,conference,anonymous,a4paper,10pt,times]{IEEEtran}
\usepackage{amsmath}
\usepackage{amsfonts,amssymb}
\usepackage{cite}
\usepackage{graphicx} 
\usepackage{xcolor}

\usepackage{cryptocode}

\usepackage[noend]{algorithm, algpseudocode}
\usepackage{wasysym}
\usepackage{xspace}
\usepackage{url}[hyphens]
\usepackage[colorlinks=false,urlcolor=black]{hyperref}
\usepackage[shortlabels]{enumitem}
\usepackage{comment}
\usepackage[hyphenbreaks]{breakurl}

\usepackage{subcaption}
\usepackage{multirow}
\usepackage{tikz}
\usetikzlibrary{positioning}
\usetikzlibrary{decorations.pathreplacing}
\usetikzlibrary{arrows.meta}
\tikzset{>={Latex[width=1.5mm,length=1.5mm]}}
\usepackage{float}

\newcommand{\content}{\ensuremath{\mathfrak{s}}\xspace}
\newcommand{\mssg}{\ensuremath{\mathsf{m}}\xspace}
\newcommand{\recoveryRate}{\ensuremath{\mathsf{b}}\xspace}

\newcommand{\messageSet}{\ensuremath{\mathfrak{S}}\xspace}
\newcommand{\securityFactor}{\lambda}
\newcommand{\remainingReports}{\ensuremath{\mathsf{r}}\xspace}
\newcommand{\messageCount}{\ensuremath{\mathsf{s}}\xspace}
\newcommand{\malLabel}{\mathsf{malicious}}
\newcommand{\honLabel}{\mathsf{honest}}
\newcommand{\ppEphemeralKey}{\ensuremath{\mathsf{X}}\xspace}
\newcommand{\blindSenderToken}{\ensuremath{\mathsf{R}}\xspace}

\newcommand{\Commit}{\ensuremath{\mathsf{Commit}}\xspace}
\newcommand{\Open}{\ensuremath{\mathsf{Open}}\xspace}
\newcommand{\commitment}{\ensuremath{\mathsf{com}}\xspace}
\newcommand{\opening}{\ensuremath{\mathsf{op}}\xspace}
\newcommand{\Encrypt}{\ensuremath{\mathsf{Enc}}\xspace}
\newcommand{\Decrypt}{\ensuremath{\mathsf{Dec}}\xspace}
\newcommand{\symmetricKey}{\ensuremath{\mathsf{K}}\xspace}
\newcommand{\ciphertext}{\ensuremath{\mathsf{ct}}\xspace}
\newcommand{\ZZ}{\mathbb{Z}\xspace}
\newcommand{\Hash}{\ensuremath{\mathsf{Hash}}\xspace}
\newcommand{\Sign}{\ensuremath{\mathsf{Sign}}\xspace}
\newcommand{\Verify}{\ensuremath{\mathsf{Verify}}\xspace}
\newcommand{\AsymmetricKeyGen}{\ensuremath{\mathsf{KeyGen}}\xspace}
\newcommand{\secretKey}{\ensuremath{\mathsf{sk}}\xspace}
\newcommand{\publicKey}{\ensuremath{\mathsf{pk}}\xspace}
\newcommand{\signingKey}{\ensuremath{\mathsf{sgk}}\xspace}
\newcommand{\verificationKey}{\ensuremath{\mathsf{vk}}\xspace}
\newcommand{\epochSecretKey}{\ensuremath{\mathsf{esk}}\xspace}
\newcommand{\epochPublicKey}{\ensuremath{\mathsf{epk}}\xspace}
\newcommand{\publicParams}{\ensuremath{\mathsf{pp}}\xspace}
\newcommand{\signature}{\ensuremath\sigma\xspace}
\newcommand{\Sym}{\ensuremath{\mathsf{Sym}}\xspace}
\newcommand{\Sig}{\ensuremath{\mathsf{Sig}}\xspace}
\newcommand{\symkeygen}{\ensuremath{\mathsf{Sym.KeyGen}}\xspace}
\newcommand{\sigkeygen}{\ensuremath{\mathsf{Sig.KeyGen}}\xspace}
\newcommand{\symenc}{\ensuremath{\mathsf{Sym.Enc}}\xspace}
\newcommand{\symdec}{\ensuremath{\mathsf{Sym.Dec}}\xspace}
\newcommand{\sigsign}{\ensuremath{\mathsf{Sig.Sign}}\xspace}
\newcommand{\sigverify}{\ensuremath{\mathsf{Sig.Verify}}\xspace}
\newcommand{\Com}{\ensuremath{\mathsf{Com}}\xspace}
\newcommand{\comcommit}{\ensuremath{\mathsf{Com.Commit}}\xspace}
\newcommand{\comopen}{\ensuremath{\mathsf{Com.Open}}\xspace}

\newcommand{\handle}{\ensuremath{\mathsf{h}}\xspace}

\newcommand{\sender}{\ensuremath{\mathsf{Sender}}\xspace}
\newcommand{\senderMarker}{\ensuremath{\mathsf{s}}}
\newcommand{\recipient}{\ensuremath{\mathsf{Receiver}}\xspace}
\newcommand{\recipientMarker}{\ensuremath{\mathsf{r}}}
\newcommand{\identity}{\ensuremath{\mathsf{ID}}\xspace}
\newcommand{\accountabilityServer}{\ensuremath{\mathsf{AS}}\xspace}
\newcommand{\DB}{\ensuremath{\mathsf{DB}}\xspace}
\newcommand{\TagsDict}{\ensuremath{\mathsf{Chnls}}\xspace}

\newcommand{\keys}{\ensuremath{\mathsf{keys}}\xspace}

\newcommand{\setup}{\ensuremath{\mathsf{SetUp}}\xspace}

\newcommand{\ppSetup}{\ensuremath{\mathsf{PP.SetUp}}\xspace}
\newcommand{\ppKeyGen}{\ensuremath{\mathsf{PP.KeyGen}}\xspace}

\newcommand{\ppBlindTokVer}{\ensuremath{\mathsf{PP.BlindTokenVer}}\xspace}

\newcommand{\senderReportTokenGen}{\ensuremath{\mathsf{TagIssue}}\xspace}
\newcommand{\senderReportTokenVerify}{\ensuremath{\mathsf{TagReceive}}\xspace}
\newcommand{\tokenTagReport}{\ensuremath{\mathsf{TagReport}}\xspace}

\newcommand{\return}[1]{\textsf{\textbf{return}}~{#1}}
\newcommand{\Err}{\textsf{Err}\xspace}
\newcommand{\IfThen}[2]{\textsf{\textbf{if}}~#1,~#2}
\newcommand{\IfNotThen}[2]{\textsf{\textbf{if not}}~#1,~#2}
\newcommand{\IfNewlineThen}[2]{\textsf{\textbf{if}}~#1,\\#2}

\newcommand{\reputation}{\ensuremath{\mathsf{y}}\xspace}
\newcommand{\token}{\ensuremath{\mathsf{t}}\xspace}
\newcommand{\timestamp}{\ensuremath\tau\xspace}

\newcommand{\NIZKDLEQ}{\ensuremath{\mathsf{NIZKDLEQ}}\xspace}

\newcommand{\bonus}{\ensuremath{\mathsf{k}}\xspace}
\newcommand{\epoch}{\ensuremath{\mathsf{e}}\xspace}
\newcommand{\reportCount}{\ensuremath{\mathsf{x}}\xspace}
\newcommand{\scoreScale}{\ensuremath{\mathsf{N}}\xspace}
\newcommand{\recoveryExp}{\ensuremath{\mathsf{p}}\xspace}
\newcommand{\noise}{\ensuremath{\mathsf{N}}\xspace}
\newcommand{\noiseDistr}{\ensuremath{\mathcal{N}}\xspace}
\newcommand{\score}{\ensuremath{\mathsf{sc}}\xspace}

\newcommand{\probReport}{\ensuremath{\mathsf{p}}\xspace}
\newcommand{\probReward}{\ensuremath{\mathsf{q}}\xspace}

\newcommand{\reputationFunc}{\ensuremath{\mathsf{upd}}\xspace}

\newcommand{\rewardLabel}{\ensuremath{\mathsf{Rew}}\xspace}
 
\newcommand{\reportScale}{\ensuremath{\mathsf{d}}}

\newcommand{\optFunc}{f}
\newcommand{\adversaryStrategy}{\mathcal{S}}
\newcommand{\maxScore}{\ensuremath{\mathsf{M}}\xspace}
\newcommand{\scoreDomain}{\ensuremath{\mathcal{D}}\xspace}
\newcommand{\reputationDomain}{\ensuremath{\overline{\mathcal{D}}}\xspace}

\newcommand{\expTime}{\ensuremath{\mathsf{E}}\xspace}

\newcommand{\noiseMean}{\ensuremath{\mu}\xspace}

\newcommand{\generator}{\ensuremath{G}\xspace}
\newcommand{\nonce}{\mathsf{n}\xspace} 
\newcommand{\blinding}{\mathsf{r}\xspace} 
\newcommand{\blinded}{\ensuremath{\mathsf{Q}}\xspace}
\newcommand{\proof}{\ensuremath{\mathsf{z}}\xspace} 
\newcommand{\scalarField}{\ZZ_q^*}

\newcommand{\tokens}{\ensuremath{\mathsf{tokens}}\xspace}
\newcommand{\tags}{\ensuremath{\mathsf{tags}}\xspace}
\newcommand{\vks}{\ensuremath{\mathsf{vks}}\xspace}
\newcommand{\vkLimit}{\ensuremath{\mathsf{B}_\mathsf{vk}}\xspace}
\newcommand{\EmptyDict}{\ensuremath{\mathsf{EmptyDict}}\xspace}
\newcommand{\EmptyQueue}{\ensuremath{\mathsf{EmptyQueue}}\xspace}

\newcommand{\adv}{\mathcal{A}}
\newcommand{\OVerify}{\ensuremath{\mathsf{OVerify}}}
\newcommand{\OIssueTag}{\ensuremath{\mathsf{OIssueTag}}}
\newcommand{\OReport}{\ensuremath{\mathsf{OReport}}}

\newcommand{\OSend}{\ensuremath{\mathsf{OSend}}}

\newcommand{\oursystem}{\ensuremath{\mathsf{Sandi}}\xspace}

\DeclareMathOperator*{\argmax}{arg\,max}

\renewcommand{\paragraph}[1]{\vspace{.6em}\noindent\textbf{#1.}\hspace*{.5em}}
\newcommand{\subparagraph}[1]{\vspace{.6em}\noindent\textit{#1.}\hspace*{.5em}}

\newtheorem{theorem}{Theorem}
\newtheorem{definition}{Definition}

\title{Sandi: A System for Accountability}

\author{
    \IEEEauthorblockN{F. Bet\"{u}l Durak}
    \IEEEauthorblockA{Microsoft Research\\
    betul.durak@microsoft.com}
    \and
    \IEEEauthorblockN{Kim Laine}
    \IEEEauthorblockA{Microsoft Research\\
    kim.laine@microsoft.com}
    \and
    \IEEEauthorblockN{Simon Langowski}
    \IEEEauthorblockA{MIT\\
    slangows@mit.edu}
    \and 
    \IEEEauthorblockN{Radames Cruz Moreno}
    \IEEEauthorblockA{Microsoft Research\\
    radames.cruz@microsoft.com}
}

\begin{document}

\maketitle
\begin{abstract}
We present a system, Sandi, for creating trust through accountability. Concretely, we focus on online communication scenarios, where the communicating parties do not know each other, yet would benefit from a degree of initial trust. Sandi can be seen as a reputation system that measures bad behavior, with strong integrity protections and resistance to manipulation. Unlike most reputation systems, Sandi is entirely based on ``downvotes'' and therefore requires strong privacy guarantees to prevent retaliation. It utilizes a ticket-based reporting mechanism to limit who can report. We also prove that Sandi incentivizes good behavior in a well-defined sense.

Sandi is by design unidirectional, so that message senders have Sandi scores and receivers can report them for inappropriate communication, but it is designed to benefit both senders and receivers. Senders benefit, as receivers are more likely to react to communication with the added trust signal. Receivers benefit from seeing senders' scores, allowing them to make more informed decisions about which senders to trust.

Receivers do not need registered accounts and neither senders nor receivers need long-term keys. Sandi guarantees score integrity, communication privacy, reporter privacy to protect reporting receivers, and sender unlinkability. Sandi can be implemented on top of any communication system that allows for small binary data transfer.

\end{abstract}

\inappendixfalse
\section{Introduction}\label{sec:intro}
Emails, instant messages, and other communication regularly go unanswered when \textit{receivers} are overwhelmed by other seemingly equally (un)important communication from other \textit{senders}.
Generative AI exacerbates this problem, as the content alone is often insufficient to determine its legitimacy.

On the other hand, a sender may have an important matter to discuss with a receiver and high confidence that the receiver will care about it, \textit{e.g.}, a recruiter reaching out to a potential job candidate or a journalist trying to reach a source of information.
Yet, unless the sender and receiver already have an existing trust relationship, the sender may be unable to get appropriate attention from the receiver and the receiver may miss a valuable opportunity.

In this work we start with the following question: \textit{How can a sender share a reliable trust signal with a receiver to facilitate efficient communication, when they share little or no prior context?}

\subsection{Introducing Sandi}
\label{subsec:intro_sandi}
We present \oursystem: a system that enables senders to explicitly ``endorse'' communication channels with their respective receivers.
Both senders and receivers can send messages on an endorsed channel, but the senders explicitly declare that they take accountability for the appropriateness of their interactions, as judged by the receivers.\footnote{This can be repeated in the other direction for bilateral endorsement and accountability.}

\oursystem is based on a centralized \textit{Accountability Server} (\accountabilityServer) that manages a reputation system with a \textit{score} for each sender.
\accountabilityServer issues \emph{endorsement tags} that bundle together the sender's score\footnote{More precisely, it uses a coarsening of the sender's score so the sender cannot be trivially identified from its score. We call this coarsening the sender's \textit{reputation}.}, a description of the communication channel, and some other data.
The sender sends the endorsement tag to its receiver to establish the endorsed channel.
The receiver can later report the sender by sending the endorsement tag back to~\accountabilityServer.
The data flow (ignoring many details) between the different parties is summarized in \autoref{fig:info_flow}.
\begin{figure}
    \begin{center}
        \begin{tikzpicture}[node distance=3cm]
            \node (score) [draw, rectangle] {Accountability Server};
            \node (sender) [draw, rectangle, below left=1.1cm and 1.5cm of score] {Sender};
            \node (recipient) [draw, rectangle, below right=1.1cm and 0.7cm of score] {Receiver};
            
            \draw[->] ([xshift=7mm]sender.north west) -- node[sloped, anchor=center, above] {\footnotesize Tag request} ([xshift=5mm]score.south west);
            \draw[->] (score) -- node[sloped, anchor=center, below] {\footnotesize Endorsement tag} (sender);
            \draw[->] (sender) -- node[above] {\footnotesize Endorsement tag} (recipient);
            \draw[<-] (score) -- node[sloped, anchor=center, above] {\footnotesize Report} (recipient);
        \end{tikzpicture}
    \end{center}
    \caption{In \oursystem, a sender obtains an endorsement tag from \accountabilityServer, which it sends to a receiver. The receiver can optionally use the tag to report the sender.
    }
    \label{fig:info_flow}
\end{figure}

\oursystem progresses in \emph{epochs}, the duration of which is a configurable system parameter.
Roughly, \accountabilityServer collects all reports received for an epoch and computes an updated score for the sender.
Since the sender learns its new score, we protect the reporters' privacy with a differential privacy mechanism that hides the true number of reports~\cite{dwork2006differential,dwork2014algorithmic}.

\oursystem itself does not decide whether some communication is appropriate or inappropriate, as it may depend on context the sender and receiver share that is entirely unknowable to anyone else.
The sender's responsibility is to guess, based on prior or auxiliary information, that the receiver will not view its communication as inappropriate.
The receiver is at liberty to decide whether the sender guessed correctly.
Note that generally \accountabilityServer is external to the communication system the sender and receiver use, so it must not be able to learn the contents of any communication, nor be able to build profiles of reporting receivers.


We thus arrive at the following question: \textit{How can we build a reliable trust score with anonymous reports for a sender to share with a receiver to facilitate efficient communication, when they share little or no prior context?}

Even the best behaving senders will eventually be reported and previously poorly behaving senders might change to start behaving well.
To take this into account, the senders' scores recover over time, unless they receive more than a threshold number of reports within an epoch.
This leads to an interesting dynamic that a rational sender must consider in its approach.
In \autoref{sec:optimality}, we show that any rationally behaving sender should avoid getting ``too many'' reports within an epoch, until it has decided to stop using \oursystem.

\oursystem provides strong security and privacy guarantees that prevent manipulation and protect both senders' and receivers' privacy.
We introduce these guarantees below in \autoref{subsec:security_privact_obj}.

\oursystem is practical and scalable.
It uses only standard public-key cryptography, requires a data transfer of a few hundred bytes for the endorsement tags, and standard digital signatures for the sender's messages on its endorsed channels.
The storage costs for \accountabilityServer are small; in particular, \accountabilityServer does not need to store data proportional to the number of currently valid tags, which could be very large.
The asymmetry, where only senders need accounts, makes the system easier to adopt.

\paragraph{Mutual benefits from Sandi}
Receivers benefit from \oursystem, as they see the senders' scores and can efficiently filter out or deprioritize communication that is more likely to be inappropriate.
Along with other security signals, the \oursystem score can protect vulnerable populations from online scams and fraud.

Senders benefit as well, because a (high) \oursystem score can increase the likelihood that receivers take their communication seriously and react to it promptly.
Senders can also use these signals to adjust their behavior to be more efficient.

\paragraph{Incentivizing endorsements}
The receiver can distinguish between endorsed communication, non-endorsed communication, and communication from senders it has previously corresponded with.
Based on this, its client software can choose how to organize and prioritize incoming messages.

This way receivers can mitigate an undesired strategy, where a sender uses endorsed channels when its reputation is sufficiently high and switches to using non-endorsed channels until its reputation has recovered.

\paragraph{Whitewashing}
Whitewashing, where users easily quit and create new accounts, poses a challenge for all reputation systems and content moderation systems.
Systems that specifically focus on accountability are fundamentally undermined by whitewashing, as it prevents reputation owners from being held accountable~\cite{mirkovic2008building}.

Luckily, several practical solutions to whitewashing exist, including requiring proof of a strong identity~\cite{mirkovic2008building}, relying on an existing PKI~\cite{koutrouli2012taxonomy,xiao2016survey}, disadvantaging fresh accounts~\cite{friedman2001social}, and adding ``friction'' to the account creation process~\cite{koutrouli2012taxonomy,kozlov2020evaluating}, such as CAPTCHAs or phone-based verification.
These solutions cannot provide complete immunity to whitewashing, but they aim to make it too costly to be practical.

Whitewashing is an independent problem that we are not trying to solve with \oursystem.
Thus, from now on we assume that the account creation process utilizes known and commonly used whitewashing prevention techniques to limit each sender to a single or a small number of accounts.

\paragraph{Malicious reporting}
A malicious group of receivers can try to attack a sender by systematically reporting each endorsement tag from the sender.
For example, a company may be targeted by the employees of another competing company reporting each endorsement tag they receive.
However, every reputation system is vulnerable to such ``false'' reports, as they are indistinguishable from valid reports.
On the flip side, the \oursystem reputation system empowers each individual receiver to report based on their own subjective view of the sender's behavior, which cannot be achieved by a centralized content moderator or a shared set of policies.

If the sender considers this to be a concern, it must be more judicious in choosing which communication channels it wants to endorse and ensure it has sufficient reason to believe its receivers behave rationally (formalized in \autoref{def:receivers-rationality}).
It should also be noted that the reporter privacy guarantees of \oursystem apply only to individual receivers; concerted reporting behavior of multiple receivers can be detected by the sender.

\subsection{Our Contributions} \label{subsec:our_contributions}
\autoref{sec:protocol} and \autoref{sec:anonymous_reporting} provide the construction of \oursystem, including protocols and algorithms for \accountabilityServer, senders, and receivers.
\accountabilityServer maintains accounts and scores for senders, whereas receivers do not need accounts.
Reporting of inappropriate communication is done in a decentralized manner by the receivers.
The ``punishment'' for poor behavior is enforced by the receivers, who can choose to deprioritize communication from low-reputation senders.
We provide communication privacy even when receivers report the sender.

At the end of each epoch, senders' scores are updated based on the number of reports they have received.
A reporter privacy guarantee mitigates problems from coercion and retaliation, and a score integrity guarantee assures senders that their score is computed correctly.

In \autoref{def:score_function}, we introduce an abstract definition of a score function that governs how the senders' reputation evolves.
In \autoref{sec:optimality}, we use this definition to show that a rational sender (\autoref{def:senders-rationality}) will avoid getting ``too many'' reports per epoch.
We make only mild and natural rationality assumptions on the receivers (\autoref{def:receivers-rationality}). 

\subsection{Security and Privacy} \label{subsec:security_privact_obj}
\oursystem needs to provide strong security and privacy properties to avoid causing unintended harms for its users.
Any participant in \oursystem can be malicious or the target of others.

We start with the assumptions we make. We assume that there is a reliable PKI for \accountabilityServer to be verified through its public key.
We assume that \accountabilityServer and senders do not collude: if a malicious \accountabilityServer colludes with a malicious sender, \accountabilityServer can manipulate the sender's score freely.
In this case, they can also break any notions of reporter's privacy without being caught by keeping track of which tag was sent to which receiver, or even just by matching timestamp data.
We argue that \accountabilityServer undermines its own utility by colluding with senders. 
In \autoref{subsec:colluding_as_and_sender}, we discuss how to technically prevent such collusion.

There are two well-known ways to limit the impact of collusion: (cryptographic) Multi-Party Computation (MPC) and (hardware-based) Trusted-Execution Environments (TEEs).
With either of these techniques, \accountabilityServer has limited or no capability to deviate from the protocol execution without being detected.
As yet another layer of protection, our design can be extended to provide ``report transparency'' for reporters to verify that their reports were correctly included in the score computations.
We leave it as a future work.



Informally, \oursystem provides the following security and privacy objectives:
\begin{itemize}
\item There could be a few ways an honest sender becomes a victim of slandering, where the system is manipulated to artificially reduce the sender's score.
(1) A malicious \accountabilityServer could try to incorrectly lower an honest sender's score without being caught. 
(2) A malicious receiver could try to report an honest sender more times than the number of valid endorsement tags it receives from the honest sender.
(3) A malicious sender could try to create endorsement tags associated with an honest sender.

\vspace{0.1cm}\noindent
\textit{Score Integrity.} We address these issues all together as follows.
One of the senders, the victim, is honest, while the rest of the senders, \accountabilityServer, and some subset of receivers are all malicious. 
Let $\messageCount$ be the number of endorsement tags the honest sender sends. 
Some of these tags are sent to malicious receivers and some to honest receivers: $\messageCount = \messageCount_\malLabel + \messageCount_\honLabel$. 
Let $\reportCount_\honLabel$ be the number of reports from honest receivers. 
Without loss of generality, we assume that all tags sent to malicious receivers are reported. 
The adversary's goal is to make an honest sender accept that the total number of reports it received is greater than $\reportCount_\honLabel + \messageCount_\malLabel$.
The game in \autoref{fig:repIntegrity} captures the idea that this is impossible.

\item A malicious \accountabilityServer could try to find who a sender communicates with and the content of the communication.

\vspace{0.1cm}\noindent
\textit{Communication Privacy.} If an honest sender establishes an endorsed channel with an honest receiver, a malicious \accountabilityServer cannot learn any information about the receiver.
Further communication on the endorsed channel remains entirely between the sender and receiver.
However, \accountabilityServer learns (1) whether a report is made, and (2) information about the reporter that is leaked by network traffic, unless reporting is done through anonymous routing.

\item A malicious sender could try to detect if a receiver reported the sender to an honest \accountabilityServer.

\vspace{0.1cm}\noindent
\textit{Reporter Privacy.}
We protect against this by hiding the true number of reports with a differential privacy guarantee.
The ensuing privacy notion is subtle; we discuss it in detail in \autoref{sec:anonymous_reporting} and \autoref{subsec:extended_privacy_discussion}.

\item A malicious receiver could try to distinguish if two distinct sender addresses belong to the same owner based on their endorsement tags.

\vspace{0.1cm}\noindent
\textit{Unlinkability.} A malicious receiver has no advantage in guessing whether two endorsement tags come from two different honest senders or the same sender beyond what can be inferred from the senders' addresses used to send the endorsement tags, content of further communication, timestamps, network traffic, and senders' reputations, without colluding with a malicious \accountabilityServer.
We formalize this security notion in \autoref{fig:realWorldSenderUnlink} and \autoref{fig:idealWorldSenderUnlink}.
\end{itemize}

\section{Related work} \label{sec:related_work}
\oursystem has similarities with works in three distinct areas: content moderation, reputation systems, and spam filtering techniques.
We compare our work with content moderation in-depth here.
We provide a summary for reputation systems and spam filetring here and refer the reader to \autoref{subsec:reputation_systems} and \autoref{subsec:spam_filters} for a much more in-depth discussion.

\subsection{Content Moderation}\label{subsec:content_moderation}
Content moderation~\cite{kamara2022outside} is an umbrella term that refers to a large number of techniques to enforce policies within communication systems.
The \emph{message franking} technique~\cite{kamara2022outside,meta_message_franking,tyagi2019asymmetric,issa2022hecate} has received particularly much attention recently, as it can enable content moderation in end-to-end encrypted messaging systems with limited violations of communication privacy.
The differences to \oursystem are significant and complex, as we will explain below.

\paragraph{Conceptual differences}
Content moderation works within policies set by the communication service.
A moderator, which may be the same as the communication service, analyzes messages and notifies the communication service about policy violations.
In end-to-end encrypted applications, receivers must choose to reveal the message to the moderator.
The beneficiary of content moderation is the communication service itself.
Senders do not benefit, but cannot opt out of content moderation either.

In contrast, \oursystem provides a trust signal that benefits senders and receivers that share little or no prior context.
Senders gain credibility and visibility by staking their reputation, and receivers can make more informed decisions based on this additional trust signal.
Report decisions in \oursystem are not bound to a specific policy and there is no moderator to make such decisions.
Instead, senders must determine whether they understand the receiver's context sufficiently well to stake their reputation.
The receiver then has the power to judge the conversation with the sender within their own context and hold the sender accountable if the conversation appeared inappropriate.
Reports have a clear and direct consequence: the sender's score changes according to a very specific kind of score function.

\paragraph{Message forwarding}
Different content moderation systems treat forwarded messages differently.
For example, the asymmetric message franking system Hecate~\cite{issa2022hecate} has an in-built traceback mechanism that allows the moderator to know who sent the message in the first place, potentially protecting the identities of intermediate receivers.
However, this decision inevitably leads to context collapse problems~\cite{marwick2011tweet}, where the forwarded receivers misinterpret the message as it is now removed from the shared context between the original sender and receiver.

In \oursystem, we make the simple decision to bind endorsement tags to the original receiver's address, automatically invalidating them when they are forwarded to other receivers.
This is a natural decision to us, as the sender voluntarily chose to endorse the communication channel with the particular receiver, indicating that they have sufficient shared context with that particular receiver---not necessarily with others.
On the other hand, a forwarded endorsement tag can very well be reported by the new receiver, because our reporting protocol does not authenticate the reporting receiver.
In any case, we ensure that each endorsement tag can be reported only once.

\paragraph{Privacy guarantees}
As long as the communication service's policies need to be interpreted by humans, full communication privacy seems impossible to achieve.
While a part of a content moderator's job may be possible to automate, there must always be an option for a human to look at the message, evaluate the context, interpret the policy, and make a final determination of whether the policy was violated.

Content moderation system specifications tend to leave open the consequences for policy-violations, which may leak an arbitrary amount of information about the reporting receiver back to the sender.
For example, the moderator may show the message to the sender as a proof of a report and of a violation, making report privacy impossible.
If no such proof is provided to the sender, a misbehaving moderator can forge evidence against an innocent sender.

In contrast, \oursystem ensures full communication privacy: endorsed channels messages communicated on them always remain between the sender and the receiver.
Still, our design allows \accountabilityServer to provide senders with irrefutable evidence of reports and proof of correct score updates.
\oursystem also provides a formal report privacy guarantee, protecting reporting receivers from coercion and retaliation.

\paragraph{Differences between Sandi and Hecate}
Despite some similarities, Hecate~\cite{issa2022hecate} and \oursystem are fundamentally different. 
Hecate reveals messages to the moderator, \oursystem provides a crowd-sourced reputation without such leakage. 
\oursystem provides (score) transparency and reporter privacy guarantees, neither of which Hecate has, by using anonymous tokens. 
This is non-trivial and must be done carefully to maintain sender unlinkability.

\oursystem is an epoch-based system, which makes tag validity times extremely tricky to handle, as we discuss in \autoref{subsec:delayed_reporting} and \autoref{subsec:receiver-setup}.
Hecate does not need to progress in epochs, because they do not provide a privacy guarantee for reporters.
Therefore, they do not need to solve similar timing problems, which makes their design much simpler than ours.

As Hecate, \oursystem also uses ephemeral keys, but they serve a completely different purpose.
Importantly, the ephemeral secret keys are only known to the senders.
At the end of each epoch, \accountabilityServer shows evidence to the senders that it has received some number of valid reports.
Senders accept this evidence, because forgery would require knowledge of the ephemeral secret keys.
With this design, \accountabilityServer never needs to store data proportional to the number of currently valid tags.

Our endorsement tag follows a similar structure as in Hecate, including a commitment to the message, an encrypted sender ID, and a timestamp.
In Hecate the tags are signed by the moderator, whereas in \oursystem they are signed by \accountabilityServer.
One advantage of Hecate is that the moderator can issue tags to the senders ahead of time; however, the communication system still signs all passing messages along with fresh timestamps.
\oursystem can be modified to do this as well, but doing this is less meaningful for us since our tags are not per message but per a specific time duration.

Message franking systems must provide some form of deniability guarantee, which Hecate achieves using moderator-generated ephemeral keys. 
Since the ephemeral public key is attached to each ``frank'', receivers can tell whether two messages originate from the same sender, inevitably violating unlinkability.

\paragraph{Comparison to FACTS}
FACTS~\cite{facts2022} is a content moderation proposal that was designed to fight fake news.
It provides a threshold approach to complaint tallying: if sufficiently many users reported a message as fake news, then the contents of the message are revealed to a moderator.
The insight in this is that fake news, by definition, are viral, and hence received and seen by many users, making it likely that reports for a message containing fake news will pass the threshold.

The technical approach in FACTS is entirely different from ours, but their goal is to some extent similar, with the exception that they break communication privacy (to some extent) and reporter privacy once the reporting threshold is reached, whereas we maintain these in any case.
They also guarantee a form of integrity for the report counts, not entirely unlike our score integrity guarantee.

\subsection{Reputation systems}
There exists a vast amount of literature on reputation systems.
Good surveys can be found in \cite{josang2007survey,hendrikx2015reputation}.

Particularly relevant to \oursystem are privacy-preserving reputation systems~\cite{gurtler2021sok,hasan2022privacy}, where various techniques are used to limit participants' visibility into some aspects of the system.
Schiffner \textit{et al.}~\cite{schiffner2011limits} and Clau{\ss} \textit{et al.}~\cite{clauss2013k} present formal privacy definitions for reputation systems, but their setting differs too much from ours for a meaningful comparison.

We defer further discussions to \autoref{subsec:reputation_systems}.

\subsection{Spam filtering}
Traditional spam filters operate on a per-message basis and offer limited user agency, whereas \oursystem gives receivers a trust signal for the entire communication channel.
Unlike traditional spam filters, \oursystem allows senders to view their scores so they can strive to maintain a good reputation, thus incentivizing good behavior.

Unlike spam filtering techniques~\cite{ramachandran2007filtering,grier2010spam,esquivel2010effectiveness,chirita2005mailrank,prakash2005fighting,zheleva2008trusting,zhang2009ipgrouprep,guzella2009review,dada2019machine}, the aim of \oursystem is not to block anything, but empower senders by allowing them to append an explicit endorsement to their communications, and empower receivers to hold senders accountable through the reporting system.
The trust signal \oursystem provides stems from this empowerment for direct action, rather than from an obscure mechanism classifying particular messages as bad.

That said, \oursystem can complement traditional spam filters, for example, by using the sender's \oursystem score as an additional signal for the spam filter, or by using a spam filter's classification as a trigger for an automated \oursystem report.
In contrast to spam filters, \oursystem senders know their standing with the system.
The receivers' view is closer to ``report spam'' buttons common in email client software today.

We defer further discussions to \autoref{subsec:spam_filters}.

\section{Overview}
\subsection{Setup}
The parties involved are the accountability server \accountabilityServer, a set of senders, and a set of receivers.
Each sender must register with~\accountabilityServer.
In this process, \accountabilityServer assigns the sender an internal account identity $\identity_\senderMarker$ and creates a corresponding database record for it.
This record includes the sender's score \score and a report count \reportCount initialized to zero.

Both \accountabilityServer and the sender can see the sender's score (a real number), whereas receivers see only a coarsening of the score, which we call the sender's \textit{reputation}~$\reputation_\senderMarker$.
The reputation is an element of some small totally ordered set, for example \{``low'', ``medium'', ``high'', ``very high''\}.
It is computed from the score through some order-preserving reputation function~$\reputation(\cdot)$.
This prevents the receivers from trivially identifying the sender by its score, as this would break unlinkability (\autoref{subsec:security_privact_obj}).
Full details are in \autoref{subsec:score-and-reputation}.

Time in \oursystem proceeds in fixed-length epochs, \textit{e.g.}, a day, a week, or a month.
During each epoch, a set of endorsement tags are sent by the senders to their desired receivers, establishing endorsed channels.
Each receiver may then report the sender by forwarding the tag to~\accountabilityServer.
At the end of the epoch, a score function \reputationFunc is used to compute the sender's score for the next epoch from its current score and the number of reports.
Finally, the report counter \reportCount is reset to zero.

\subsection{Building up to Sandi}
\label{sec:overview-building-up}
\paragraph{First attempt}
We start from a construction technically similar to the asymmetric message franking scheme Hecate~\cite{issa2022hecate}.
For more details about the relationship between \oursystem and message franking, including an explanation of the differences to Hecate, see \autoref{subsec:content_moderation}.

When a sender wants to send an endorsement tag to a receiver, it needs to first authenticate with \accountabilityServer so that \accountabilityServer knows the sender's account $\identity_\senderMarker$.
The sender chooses a short-term signing key $\signingKey_\senderMarker$ and a corresponding verification key~$\verificationKey_\senderMarker$; later it uses $\signingKey_\senderMarker$ to sign its communication on the endorsed channel.
It takes care not to use the same signature key pair for endorsing communication from multiple addresses.
Next, it computes cryptographic commitments $(\commitment_\senderMarker, \opening_\senderMarker)$ and $(\commitment_\recipientMarker, \opening_\recipientMarker)$ to $\verificationKey_\senderMarker$ and the receiver's address $\handle_{\recipientMarker}$, respectively.
It sends both commitments to~\accountabilityServer.

Subsequently, \accountabilityServer encrypts the sender's $\identity_\senderMarker$ with a secret encryption key \symmetricKey as $\ciphertext \gets \Encrypt_\symmetricKey(\identity_\senderMarker)$ and uses a secret signing key \secretKey to produce a digital signature $\signature \gets \Sign_{\secretKey}(\commitment_\senderMarker||\commitment_\recipientMarker||\timestamp||\reputation_\senderMarker||\ciphertext)$.
Here \timestamp is a timestamp and $\reputation_\senderMarker$ the sender's reputation.
It creates an endorsement tag $\token \gets (\commitment_\senderMarker, \commitment_\recipientMarker, \timestamp, \reputation_\senderMarker, \ciphertext, \signature)$, which it sends to the sender.

Once the sender has received the endorsement tag, it verifies that the signature is valid using \accountabilityServer's public verification key.
It then sends the endorsement tag $(\opening_\senderMarker, \opening_\recipientMarker, \verificationKey_\senderMarker, \token)$ to the receiver, thus establishing an endorsed channel.
The sender uses $\signingKey_\senderMarker$ to sign further communication on this channel.

Upon receiving and parsing $(\opening_\senderMarker, \opening_\recipientMarker, \verificationKey_\senderMarker, \token)$, the receiver opens the commitments and verifies that \signature is valid.
It verifies the signatures on any further communication from the sender on this channel using~$\verificationKey_\senderMarker$.
It uses the sender's reputation $\reputation_\senderMarker$ as additional evidence to decide how to respond.
If the receiver considers the sender's communication to be inappropriate, it can report the sender by sending \token to~\accountabilityServer.

Upon receiving and parsing a report \token, \accountabilityServer checks \token has not yet been reported, verifies \signature, decrypts \ciphertext to obtain $\identity_\senderMarker$, and increments the sender's report count~\reportCount.
When the epoch changes, \accountabilityServer uses a \emph{score function} $\reputationFunc$ to update the sender's score as $\score \gets \reputationFunc(\score, \reportCount)$.

\paragraph{Communication privacy and unlinkability}
As long as the sender does not use the same signature key pair to communicate from multiple addresses, both communication privacy and unlinkability (\autoref{subsec:security_privact_obj}) are already satisfied by the above construction.
Namely, the hiding property of the commitment scheme (\autoref{appendix:crypto_prelim}) ensures that \accountabilityServer cannot learn the receiver's address from the commitment.
The unlinkability property holds, because the receiver cannot learn the sender's identity due to the symmetric encryption.

\paragraph{Score integrity with malicious AS} 
The construction above illustrates the core idea behind \oursystem, but it lacks a strong score integrity guarantee (\autoref{subsec:security_privact_obj}).
Namely, we want to ensure senders that \accountabilityServer is not misbehaving to reduce their scores arbitrarily.

We do this by incorporating a Privacy Pass~\cite{davidson2018privacy} (\autoref{appendix:crypto_prelim}) blind token mechanism into the endorsement tag request-report flow, with senders working as issuers and \accountabilityServer as the redeemer.
The idea is that with each endorsement tag the sender attaches a (blind) Privacy Pass token that \accountabilityServer will be able to get only if the receiver shares it as a part of a report.
\accountabilityServer can subsequently unblind these tokens to obtain valid Privacy Pass tokens; we call these additional tokens \emph{sender-tokens}.
To prove it has received reports, \accountabilityServer shows the sender-tokens to the sender, who can verify their validity with an epoch-ephemeral secret key.

We note that the sender-tokens are unlinkable from the endorsement tags in the sense that the sender cannot tell which sender-token corresponds to which endorsement tag.

\paragraph{Reporter privacy}
It is possible for a sender to learn whether a particular receiver reports its endorsement tag.
In an extreme case, the sender may send out only a single endorsement tag, in which case the number of reports immediately reveals whether the receiver reported it.
For example, a receiver may not want to report a business partner's inappropriate communication to avoid burning bridges or risking escalation.
As another example, a receiver may not want to report a spammer's endorsement tag to not inform the spammer that its communication went through to a human receiver.

We address this using differential privacy~\cite{dwork2006differential,dwork2014algorithmic}.
At the end of each epoch, \accountabilityServer samples an integer $\noise \gets \noiseDistr$ from some appropriate noise distribution $\noiseDistr$ and silently drops $\noise$ reports for the sender.
This can be turned into a formal guarantee that the sender cannot learn if any particular receiver reported it.

\subsection{Optimality}\label{subsec:optimality}
We argue that \oursystem incentivizes senders to behave ``not too badly'', until they stop using \oursystem.
This guarantee is not trivial.
For example, a sender may want to set up several endorsed channels for malicious and abusive---yet profitable---communication. 
Why would the sender not just go ahead with its operation and stop using \oursystem once its score crashes?
How can we possibly prevent such behavior?

One solution would be to let \accountabilityServer decline issuing the sender endorsement tags when its score is low.
However, if the sender has a fresh account with a high score, \accountabilityServer cannot know the sender's intentions since it cannot see the sender's communication.
We take another approach, where we incentivize ``good'' behavior by making it more advantageous for senders to delay bad behavior until arbitrarily much later.

In \autoref{sec:optimality} we analyze a rational sender (\autoref{def:senders-rationality}) trying to maximize its benefit.
We prove that every rational sender has an optimal strategy of a specific form (\autoref{th:opt_strategy_1}) that incentivizes it to engage in (mostly) appropriate communication for as long as it keeps using \oursystem, and that this incentive can be controlled by adjusting~\reputationFunc (\autoref{th:opt_strategy_2}).

\section{Preliminaries}

\subsection{Scripts and Rationality}\label{subsec:conversations-and-scripts}

\paragraph{Scripts} In \oursystem, time is divided into epochs.
During each epoch, senders request and send out a set of endorsement tags for specific communication channels.
Each endorsement tag encodes the sender's reputation $\reputation_\senderMarker$ and is signed by~\accountabilityServer.
For now, suppose endorsement tags and endorsed channels remain valid until the end of the epoch in which they were created.

Once the sender and receiver have established an endorsed channel, they use it for bidirectional communication.
The receiver can then report or reward the sender (or do both) based on the communication and the sender's reputation.
Reporting works as we outlined in \autoref{sec:overview-building-up}, whereas rewarding represents an event where the sender benefits from the communication in some measurable way. 

To study the probabilities of these events and the expected rewards, the sender models its behavior as a \emph{script} \content that it executes (think of a sales script).
The script determines the sender's strategy in choosing its own messages, but is independent of the receiver's (realized) choices.
We denote the space of all scripts that the sender is willing to execute at epoch index $i$ by $\messageSet_i$.
Since different scripts work differently for different receivers, we encode the receiver's address in every script.
Thus, a script $\content \in \messageSet_i$ is a random variable that, when ``executed'', outputs a full communication between the sender and receiver.
From now on we denote $\messageSet_i$ by just~$\messageSet$, as the epoch index is always obvious from the context.

Senders cannot influence the evolution of \messageSet through their actions.
For example, executing a script does not cause another script to be included in or excluded from \messageSet later.

We denote by $\probReward(\content, \reputation_\senderMarker)$ and $\probReport(\content, \reputation_\senderMarker)$ random variables that output the reward and report probabilities when the sender with reputation $\reputation_\senderMarker$ executes a script~\content.
Similarly, we denote by $\rewardLabel(\content)$ a random variable that outputs the sender's reward.
The sender can empirically estimate the expectation values for these random variables for any script it is willing to execute.\footnote{It is unrealistic for the sender to study these expectation values for each receiver independently.
In practice, the sender may know some properties of a receiver, \textit{e.g.}, their age bracket, their income bracket, or their interests.
It then uses marginal expectation instead, integrating over subsets of similar receivers.
This has no impact on our analysis.}
We formalize these notions as follows:
\begin{definition}[Reward function]\label{def:reward_function}
The reward function $\rewardLabel(\cdot)$ maps a script $\content \in \messageSet$ to a random variable $\rewardLabel(\content)$ with range~$(0, \infty)$.
\end{definition}
\begin{definition}[Reward probability function]\label{def:reward_probability}
For any script $\content \in \messageSet$, the reward probability function $\probReward(\content, \cdot)$ maps a sender's reputation $\reputation_\senderMarker$ to a random variable $\probReward(\content, \reputation_\senderMarker)$ with range~$[0, 1]$.
\end{definition}
\begin{definition}[Report probability function]\label{def:report_probability}
For any script $\content \in \messageSet$, the report probability function $\probReport(\content, \cdot)$ maps a sender's reputation $\reputation_\senderMarker$ to a random variable $\probReport(\content, \reputation_\senderMarker)$ with range~$(0, 1]$.\footnote{Excluding 0 rules out a trivial strategy, where the sender may be able start a large number of endorsed channels with a zero chance of being reported, yet each yielding a positive expected reward.
This is a realistic assumption that can be enforced by making the receiver software automatically report any endorsement tag at random with a negligible probability.}
\end{definition}
We introduce the notation
\begin{equation} \widetilde{\rewardLabel}(\content, \reputation_\senderMarker) := \frac{\mathbb{E}[\probReward(\content, \reputation_\senderMarker) \rewardLabel(\content)]}{\mathbb{E}[\probReport(\content, \reputation_\senderMarker)]} \in [0,\infty) \label{eq:rew-tilde-definition} \end{equation}
that will prove to be very convenient in \autoref{sec:optimality}.

We can now also define what rational behavior means for receivers and senders:
\begin{definition}[Rational receiver]\label{def:receivers-rationality}
We say that a receiver is rational if, for any $\content \in \messageSet$, (1) $\mathbb{E}[\probReward(\content, \cdot) \rewardLabel(\content)]$ is an increasing function (not necessarily strictly increasing), (2) $\mathbb{E}[\probReport(\content, \cdot)]$ is a decreasing function (not necessarily strictly decreasing), and (3) $\probReward(\content, \reputation_\senderMarker) \rewardLabel(\content)$ and $\probReport(\content, \reputation_\senderMarker)$ are independent for any~$\reputation_\senderMarker$.\footnote{The third condition is natural, because the receiver can independently both reward and report the sender.}
\end{definition}
\begin{definition}[Rational sender]\label{def:senders-rationality}
We say that a sender is rational if it behaves to ``maximize its expected total reward''\footnote{We postpone formalizing what ``maximize its expected total reward'' means to \autoref{def:optimal_senders_strategy} in \autoref{sec:optimality}. Despite \autoref{def:senders-rationality} being slightly incomplete for now, it gives a good idea of the kind of assumption we are working with.} throughout its use of \oursystem and only executes scripts \content for which it knows $\mathbb{E}[\probReward(\content, \cdot) \rewardLabel(\content)]$ and $\mathbb{E}[\probReport(\content, \cdot)]$.
\end{definition}
For example, a rational sender working for a company would never set up an endorsed channel with a receiver from another competing company without believing that the receiver will not systematically report its endorsement tags.

We must emphasize that \autoref{def:receivers-rationality} and \autoref{def:senders-rationality} are not necessary for any of the security and privacy guarantees of \oursystem.
Indeed, since receivers' privacy is strongly protected, nothing prevents them from behaving completely arbitrarily if they want.
In such a situation it would be impossible to say anything about how senders may want to behave, since receivers can choose to, for example, entirely ignore senders' reputation.
However, in \autoref{sec:optimality} we show that \emph{if} a sender knows (or in practice has high confidence) that its receivers behave rationally, then the sender's optimal behavior is to not receive too many reports.

We require \messageSet to satisfy a few technical properties, which we present in \autoref{sec:props-of-script-space}.

\paragraph{Examples}
What \messageSet concretely is depends on the scenario.
For example, the sender may work at the sales department of a company and be instructed to sell one of a few products following an existing sales pitch.
They may follow a set of approved strategies or deviate into aggressive and unscrupulous strategies to increase their sales; the script space for this sender consists exactly of the strategies they are willing to follow.
As another example, a recruiter may email a candidate or talk with them on the phone.
However, it is not uncommon for dishonest recruiters to mislead candidates about the true nature of the work.
In this case, the script space captures the complete strategies the recruiter is willing to use.
The ``conversation'' that thus plays out may (depending on the tag validity time) include the candidate's experience starting to work for their new employer.
It is this entire experience that the candidate can now choose to report.

It is worth emphasizing the remarkable flexibility of the approach we take here and that the optimality results in \autoref{sec:optimality} hold in all cases.
Namely, the sender's optimal strategy is in any case to behave ``not too badly'', until they decide to stop using \oursystem.

\subsection{Score and Reputation}\label{subsec:score-and-reputation}
All reputation systems use some kind of a \textit{score function} that governs how reputation evolves.
Common choices involve averages or (weighted) sums, but also many other more exotic functions have been used.
Most such naive choices for score functions result in the emergence of undesirable optimal strategies, which can be thought of a ways to ``game'' the system.
In this section we introduce our own carefully crafted scoring mechanism that allows us to prove the optimality results in \autoref{sec:optimality}, essentially eliminating such undesirable optimal strategies.

By a score function $\reputationFunc(\score, \reportCount)$ we mean a function that outputs the sender's score for the next epoch based on the reports it has received in the current epoch.
We take an axiomatic approach, listing properties that any $\reputationFunc$ must satisfy, all of which are needed for our optimality results.
\begin{definition}[Score function]
\label{def:score_function}
Let $\bonus \in \mathbb{Z}_{\geq 1}$, $\maxScore \in \mathbb{Z}_{\geq 1}$, $\reportScale \in (0, 1]$, and $\scoreDomain := (-\infty, \maxScore]$.
A score function $\reputationFunc_\bonus^{\maxScore, \reportScale}:\scoreDomain \times \mathbb{Z} \rightarrow \scoreDomain$ is a function that satisfies the following properties:
\begin{flalign} 
&\reputationFunc_\bonus^{\maxScore, \reportScale}(\score + \reportScale, \reportCount) \geq \reputationFunc_\bonus^{\maxScore, \reportScale}(\score, \reportCount - 1) \label{eq:upd_reputation_vs_report} \\
&\reputationFunc_\bonus^{\maxScore, \reportScale}(\score, \reportCount) \geq \reputationFunc_\bonus^{\maxScore, \reportScale}(\score, \reportCount + 1) \label{eq:upd_report_are_bad} \\
&\reputationFunc_\bonus^{\maxScore, \reportScale}(\score, \reportCount) \geq \reputationFunc_\bonus^{\maxScore, \reportScale}(\score, \reportCount + 1) + \reportScale\,\,\text{if}\,\,\reportCount \geq \bonus \label{eq:upd_reports_are_really_bad} \\
&\reputationFunc_\bonus^{\maxScore, \reportScale}(\score, \bonus) \geq \score \label{eq:upd_grace_score}
\end{flalign}
We often simplify notation and write $\reputationFunc := \reputationFunc_\bonus^{\maxScore, \reportScale}$.
\end{definition}
Several comments are in order.
The constant \maxScore provides an upper bound on the score values, capturing the idea of monotonically decreasing reputation.
The constant $\reportScale$ determines the impact of a single report on the score.
The definition allows for a negative \reportCount: this is a technicality and will be important in \autoref{sec:anonymous_reporting}.
The domain $\scoreDomain$ indicates that the score values can be real numbers and not just integers.
Equation \eqref{eq:upd_reputation_vs_report} balances the benefit of a higher score with the disadvantage of an additional report; \eqref{eq:upd_report_are_bad} states that more reports lead to a lower score; \eqref{eq:upd_reports_are_really_bad} states that this effect is strict when the number of reports exceeds the \emph{tolerance level}~\bonus.
Finally, \eqref{eq:upd_grace_score} ensures senders can either maintain or recover their score by getting no more than \bonus reports per epoch.

In \oursystem a sender and \accountabilityServer have a precise view of the sender's score~\score\footnote{This is not quite true: \accountabilityServer sees a more accurate non-private view of the sender's score and only shares \score with the sender through a differential privacy mechanism (\autoref{sec:anonymous_reporting}).}, but receivers are shown only a coarsening $\reputation_\senderMarker \gets \reputation(\score)$, because revealing \score would be too identifying of the sender.
We call the function $\reputation$ a \textit{reputation function}, and $\reputation_\senderMarker$ the sender's reputation.

\begin{definition}[Reputation function]
\label{def:reputation_function}
A reputation function $\reputation$ is an order-preserving function from $\scoreDomain$ to a finite non-empty totally ordered set $\reputationDomain$.
\end{definition}
$|\reputationDomain|$ must be much smaller than the number of senders to prevent senders from being identified by their reputation.
For example, \reputationDomain may be \{``low'', ``medium'', ``high'', ``very high''\}.

\paragraph{Example score function family}
Let $\recoveryRate \in (0,\reportScale]$. Then the following function satisfies the requirements in \autoref{def:score_function} for $\reputationFunc_\bonus^{\maxScore, \reportScale}$:
\begin{equation}\small
  \reputationFunc(\score, \reportCount) :=
    \begin{cases}
      \score - \reportScale(\reportCount - \bonus), & \reportCount \geq \bonus \\
      \min  \left\{ \score + \recoveryRate, \maxScore \right \}, & \reportCount < \bonus,\, \score \geq 0 \\
      \min \left\{ \score + \reportScale(\bonus - \reportCount), 0 \right \}, & \reportCount < \bonus,\, \score < 0\\
    \end{cases}
    \label{eq:our_upd}
\end{equation}

A generalization of the function family in \eqref{eq:our_upd} is presented in \autoref{app:generalized_score_func}, along with a proof that it satisfies \autoref{def:score_function}.

\section{Construction of Sandi}
\label{sec:protocol}
In this section we present the construction and protocol interface for \oursystem.
The construction here satisfies all of the security and privacy properties in \autoref{subsec:security_privact_obj} except reporter privacy.
We extend to cover reporter privacy in \autoref{sec:anonymous_reporting}.

\subsection{Communication Channels}\label{subsec:comm_channels}
As we mentioned in \autoref{subsec:intro_sandi}, we assume the account creation process is Sybil-resistant in the sense that a sender entity cannot register more than one account with~\accountabilityServer.
Thus, the senders and accounts are in a one-to-one correspondence.

The idea in \oursystem is that the sender sends an endorsement tag through some communication channel from its own address to a receiver's address that we denote~$\handle_\recipientMarker$.
The sender includes with the endorsement tag a short-term signature verification key $\verificationKey_\senderMarker$ and keeps the corresponding signing key $\signingKey_\senderMarker$ to itself.
The endorsement tag establishes a unilateral (sender's) endorsement for the channel; the channel can then be used for back-and-forth communication.
When the sender sends a message \mssg on the channel, it includes a signature $\Sign_{\signingKey_\senderMarker}(\handle_\recipientMarker|| \mssg)$ binding the message to the endorsed channel.\footnote{Signing communication on the endorsed channel prevents the channel from being spoofed by another sender and prevents message tampering.
It also protects the sender if its endorsement tag is stolen, as long as it keeps the signing key safe.}
The endorsement remains valid for a limited time, after which a new endorsement tag must be sent.

\begin{definition}[Communication channel]
A communication channel is a pair $(\verificationKey_\senderMarker, \handle_\recipientMarker)$ consisting of a sender's signature verification key and a receiver's address.
\end{definition}
Generally, senders and receivers can control multiple different addresses, \textit{e.g.}, email addresses or various instant messaging handles.
However, to maintain its unlinkability guarantee, the sender must never use a single signing key pair with multiple addresses.
Otherwise, a receiver can identify that two tags with the same $\verificationKey_\senderMarker$ belong to the same sender, even if they come from two apparently unrelated sender addresses.

The sender can in principle change its signature key pair for each new endorsement tag; hence the signature key pair is inherently short-term.
In more detail, when requesting a new tag from \accountabilityServer, the sender includes a hiding and binding (reusable) commitment (\autoref{appendix:crypto_prelim}) to the verification key it intends to use.
It sends this commitment to \accountabilityServer as a part of the tag request, enabling \accountabilityServer to maintain a count (upper bound) of the number of keys concurrently in use by this sender. 
This limitation is important for our reporter privacy guarantee, as we explain in \autoref{sec:anonymous_reporting}.

\subsection{Tag Expiration and Delayed Reporting}\label{subsec:delayed_reporting}
\oursystem uses two different notions of expiration time for endorsement tags: \textit{expiration time for reporting} and \textit{expiration time for receiving}. 

To prevent slandering attacks, where a malicious receiver collects and reports multiple tags all at once, we require \accountabilityServer to reject tags that were issued too far in the past.
For this, we define the tag \textit{expiration time for reporting} as
\[ \mathsf{exp\_time\_for\_reporting} := \mathsf{iss\_time} + \expTime\cdot \mathsf{epoch\_dur}, \]
where $\mathsf{iss\_time}$ denotes the tag's time of issuance, $\mathsf{epoch\_dur}$ denotes the duration of an epoch, and $\expTime \geq 0$ is an integer.

On the other hand, if endorsement tags expire too quickly, receivers may not be able to realistically report them, and it would not make sense for the receiver's client program (user interface) to show them as valid.
For this reason, we require that a receiver accepts an endorsement tag as valid only if, upon first seeing it, the tag is no older than its \textit{validity period}, 
\[ \mathsf{val\_period} \leq (\expTime - 1) \cdot \mathsf{ epoch\_dur}\,. \]
Therefore, the tag \textit{expiration time for receiving} is
\[ \mathsf{exp\_time\_for\_receiving} := \mathsf{iss\_time} + \mathsf{val\_period} \,. \]
Clearly $\expTime = 0$ does not make sense in reality, as it would require that receivers can immediately receive and report messages.
If $\expTime = 1$, our desired upper bound for $\mathsf{val\_period}$ above leaves no time for the receiver to report.
Thus, we require $\expTime \geq 2$, leaving receivers with at least a full epoch duration to send a report.

In each epoch \accountabilityServer can receive reports for tags sent in the current epoch or any of the past \expTime epochs.
\accountabilityServer keeps separate counts of these, and uses only those issued \expTime epochs back to compute the senders' scores for the next epoch.
We call this idea \emph{delayed reporting}.
\autoref{fig:report_scheduling} illustrates delayed reporting in the case $\expTime = 3$.
\begin{figure}
\begin{center}
\begin{tikzpicture}
    \draw[->, thick] (0,0) -- (7,0);
    \node[above] at (7, 0) {time};

    \foreach \x/\l in {1/\timestamp, 2/\timestamp+1, 3/\timestamp+2, 4/\timestamp+3, 5/{} } {
        \draw[thick] (\x,-0.1) -- (\x,0.1);
    }

    \draw[decorate, decoration={brace, amplitude=10pt, raise=6pt}, thick] (2,0) -- (1,0);
    \node at (1.5,-0.9) {Issuance epoch};

    \draw[dashed] (5,-1.2) -- (5,1.2);
    \node[right] at (5,-1.2) {Score update};

    \draw[-, very thick, color=black] (4.1,0.4) -- (4.1,0.6);
    \draw[-, very thick, color=black] (4.2,0.6) -- (4.2,0.8);
    \draw[-, very thick, color=red] (4.6,0.8) -- (4.6,1.0);
    \draw[-, very thick, color=black] (4.8,1.0) -- (4.8,1.2);
    
    \draw[<->, thick, color=black] (1.1,0.5) -- (2.0,0.5);
    \draw[dotted, thick, color=black] (2.0,0.5) -- (4.1,0.5);
    
    \draw[<->, thick, color=black] (1.2,0.7) -- (1.8,0.7);
    \draw[dotted, thick, color=black] (1.8,0.7) -- (4.2,0.7);
    
    \draw[<->, thick, color=red] (1.6,0.9) -- (4.2,0.9);
    \draw[dotted, thick, color=red] (4.2,0.9) -- (4.6,0.9);
    
    \draw[<->, thick, color=black] (1.8,1.1) -- (2.3,1.1);
    \draw[dotted, thick, color=black] (2.3,1.1) -- (4.8,1.1);
\end{tikzpicture}
\end{center}
\caption{Delayed reporting with $\expTime = 3$. The horizontal arrows indicate the delay from tag issuance to a receiver receiving the tag. The dotted horizontal lines indicate the remaining time to report until the tag expires (vertical bar). The ``Score update'' line shows when \accountabilityServer collects the reports from the ``Issuance epoch'' to update the sender's score.
The \textcolor{red}{second tag from the top} is considered invalid by the receiver, as it was received too late (after its $\mathsf{val\_period}$).}
\label{fig:report_scheduling}
\end{figure}

It suffices for \accountabilityServer to include a single timestamp $\timestamp$ in each endorsement tag indicating the tag's issuance time, as this allows every participant to compute the corresponding expiration times as long as they know $\expTime$, $\mathsf{epoch\_dur}$, and $\mathsf{val\_period}$.

Using timestamps avoids having to assume synchronization on the notion of epochs. 
Using an epoch index instead would be impractical, as senders and receivers would need to know when the epoch changes take place.
Instead, \accountabilityServer splits time into epochs by itself and publishes the length of the epoch ($\mathsf{epoch\_dur}$).
\accountabilityServer knows how to compute an epoch index from any timestamp~\timestamp, so it can perform timestamp-to-epoch conversions internally.
Specifically, \accountabilityServer can index into an epoch-specific part of its internal database based on~\timestamp to access delayed reporting data.

\subsection{Receiver's Setup} \label{subsec:receiver-setup}
To simplify the notation, we assume that each receiver only uses one address $\handle_\recipientMarker$ for endorsed communication.\footnote{This is not a limitation; everything in this section can be generalized to multiple receiver addresses in a straightforward and practical manner. However, there are some non-trivial implications that we discuss in \autoref{subsec:extended_privacy_discussion}.}
To keep track of endorsed channels, each receiver maintains a dictionary $\TagsDict$, mapping a sender's verification key $\verificationKey_\senderMarker$ to a pair $(\timestamp_\text{rep}, \tags)$ of information about the channel $(\verificationKey_\senderMarker, \handle_\recipientMarker)$.
The timestamp $\timestamp_\text{rep}$ indicates when the receiver can again report this channel and \tags is a queue of endorsement tags for this channel.

In more detail, the role of $\timestamp_\text{rep}$ is to keep track of a \textit{lock period}, ensuring the receiver never reports an endorsed channel more than once within any
\[ \mathsf{report\_lock} \stackrel{\text{require}}{\geq} \expTime \cdot \mathsf{epoch\_dur} \]
duration of time.
This is crucial for our reporter privacy guarantee, as we explain in \autoref{sec:anonymous_reporting}.

To report the sender on an endorsed channel $(\verificationKey_\senderMarker, \handle_\recipientMarker)$, the receiver first verifies $\TagsDict[\verificationKey_\senderMarker].\timestamp_\text{rep}$ is in the past.
If so, it pops a tag from $\TagsDict[\verificationKey_\senderMarker].\tags$ and sends it to~\accountabilityServer.
It updates $\timestamp_\text{rep}$ as we show in \autoref{subsec:protocol_interface} (\autoref{proc:report_protocol}).

When receiving messages on an endorsed channel, the receiver checks the signature on the message against $\verificationKey_\senderMarker$ and checks whether the tail of $\TagsDict[\verificationKey_\senderMarker].\tags$ holds an endorsement tag that is still valid \textit{for receiving} (\autoref{subsec:delayed_reporting}).
Otherwise, it considers the channel as no longer endorsed and may want to request a fresh tag.

The head of the queue $\TagsDict[\verificationKey_\senderMarker].\tags$ holds the oldest endorsement tag for this channel, whereas the tail holds the latest one.
As a garbage collection routine, the receiver always pops the queue head after it expires \textit{for reporting} (\autoref{subsec:delayed_reporting}).
It removes the channel entry entirely once the queue becomes empty and $\TagsDict[\verificationKey_\senderMarker].\timestamp_\text{rep}$ is in the past.

\subsection{Public Parameters}\label{subsec:public_params}
\oursystem requires a symmetric encryption scheme $\Sym$ (algorithms \symkeygen, \symenc, \symdec), a signature scheme $\Sig$ (algorithms \sigkeygen, \sigsign, \sigverify), and a hiding and binding commitment scheme \Com (algorithms \comcommit, \comopen) (see \autoref{appendix:crypto_prelim} for interfaces).
For simplicity, we write $\Encrypt := \symenc$, $\Decrypt := \symdec$, $\Sign := \sigsign$, $\Verify := \sigverify$, $\Commit := \comcommit$, and $\Open := \comopen$.

We assume all parties have access to the following set of public parameters: a security parameter $\lambda$; a generator \generator for a prime order group of order $q$, where the discrete log problem is hard; a score function $\reputationFunc$ and a reputation function~$\reputation$; the epoch duration $\mathsf{epoch\_dur}$, the expiration time multiplier $\expTime$, the validity period $\mathsf{val\_period}$, and $\mathsf{report\_lock}$ (\autoref{subsec:receiver-setup}).
Each sender is limited to using at most $\vkLimit \geq 1$ different signature verification keys $\verificationKey_\senderMarker$ within any $\mathsf{report\_lock}$ duration of time.

The time parameters above can generally vary based on the sender, but the receiver must know what these are, which we assume to be the case.
The same is true for \vkLimit; for example, for consumer accounts \vkLimit would likely be very small, like 1 or 2, whereas for commercial accounts a larger value may be more practical.

\subsection{Protocol Interface}\label{subsec:protocol_interface}
\paragraph{SetUp}
\accountabilityServer runs \setup only once before doing anything else to initialize its encryption and signature keys. It creates a symmetric key $\symmetricKey \gets \symkeygen(1^{\lambda})$ and a signature key pair $(\secretKey, \publicKey)\gets \sigkeygen(1^{\lambda})$. 
\accountabilityServer generates public parameters for the Privacy Pass blind token protocol as $\publicParams \gets \ppSetup(1^\lambda)$.
This, and the other Privacy Pass algorithms, are presented in \autoref{appendix:crypto_prelim}, following~\cite{davidson2018privacy}.
\accountabilityServer keeps $\symmetricKey$ and $\secretKey$ to itself and distributes \publicKey and \publicParams to all participants. \accountabilityServer sets up an empty database~\DB.

\paragraph{Registration}
For each \sender, \accountabilityServer defines a new account identifier $\identity_{\senderMarker}\gets_\$ \{0, 1\}^*$.
Whenever there is any communication between \sender and \accountabilityServer, we assume that \sender authenticates to prove it controls the account for~$\identity_{\senderMarker}$. 
For each epoch index $i$ (increasing with time), $\DB$ includes an epoch-dependent state
\[ \DB[\identity_{\senderMarker}, i] := (\epochPublicKey, \score, \vks, \tokens), \]
where \epochPublicKey optionally holds a public key, \score is \sender's score, \vks is a dictionary mapping \sender's short-term verification keys commitments to timestamps indicating when the commitments are no longer in use, and \tokens is a set of reports.
\accountabilityServer initializes $\DB[\identity_{\senderMarker}, i_\text{now}] \gets ( \bot, \score_\text{init}, \EmptyDict, \emptyset )$.
Here $\score_\text{init} \leq \maxScore$ denotes the starting score for \sender.

We often write a timestamp $\timestamp$ instead of the epoch index $i$ when indexing into $\DB$ (\textit{i.e.}, $\DB[\identity_{\senderMarker}, \timestamp]$), assuming \accountabilityServer automatically resolves the timestamp to the correct epoch index.
We use $\timestamp_\text{now}$ to denote a timestamp for the current time.

An obvious garbage collection mechanism cleans the data for sufficiently old epochs.

\paragraph{KeyRegistration}
\sender and \accountabilityServer run this protocol in each epoch before any endorsement tags can be issued.
First, \sender uses \ppKeyGen to sample a key pair $(\epochSecretKey_\senderMarker, \epochPublicKey_\senderMarker)$ for issuing Privacy Pass tokens.
It keeps $\epochSecretKey_\senderMarker$ to itself and sends $\epochPublicKey_\senderMarker$ to \accountabilityServer.
\accountabilityServer adds $\epochPublicKey_\senderMarker$ to \sender's state for the current epoch.
The detailed key registration protocol is given in \autoref{proc:epoch_key_reg}.

\begin{figure}[ht!]
\procedureblock[colspace=-0.2cm,colsep=0ex,codesize=\scriptsize,bodylinesep=1ex]{\small\textsf{KeyRegistration}}{
    \sender( \publicParams) \<\< \accountabilityServer(\DB, \identity_{\senderMarker}, \timestamp) \< \\ 
    \pcln (\epochSecretKey_\senderMarker, \epochPublicKey_\senderMarker) \gets \ppKeyGen(\publicParams)  \<\<\< \\[0.2cm]
    \< \hspace{-1cm}\xrightarrow{\hspace{0.7cm}\mbox{\scriptsize$\epochPublicKey_\senderMarker$}\hspace{0.7cm}} \\
    \< \< \pcln \IfNewlineThen{\DB[\identity_\senderMarker,\timestamp].\epochPublicKey \neq \bot}{\<\<\qquad\quad\return{\Err}} \< \\ 
    \< \< \pcln \DB[\identity_\senderMarker,\timestamp].\epochPublicKey \gets \epochPublicKey_\senderMarker \< \\
}
\caption{Epoch key registration protocol.}
\label{proc:epoch_key_reg}
\end{figure}

\paragraph{TagIssue}
\sender inputs a short-term verification key $\verificationKey_\senderMarker$  corresponding to the address it intends to use for sending the tag (recall \autoref{subsec:comm_channels}).
It computes commitments to $\verificationKey_\senderMarker$ and to the receiver's address~$\handle_\recipientMarker$; these commitments determine the communication channel it intends to endorse.
It stores the corresponding openings $\opening_\senderMarker$ and $\opening_\recipientMarker$, and sends the commitments $\commitment_\senderMarker$ and $\commitment_\recipientMarker$ to~\accountabilityServer.
These commitments prevent the endorsement tag from being maliciously repurposed, as otherwise an attacker may be able to steal the tag and use it to endorse its own channel with the same (real) receiver, or the (real) receiver may try to forward it to another receiver, setting up a false endorsement with the (real) sender.

\accountabilityServer inputs its secret keys, $\DB$, and $\identity_\senderMarker$.
It records the current time $\timestamp \gets \timestamp_\text{now}$.\footnote{If $\DB[\identity_\senderMarker, \timestamp]$ does not exist, \accountabilityServer needs to end the previous epoch (see below), then together with \sender run $\mathsf{KeyRegistration}$. We assume this has been done.}
It looks up \sender's key $\epochPublicKey_\senderMarker \gets \DB[\identity_\senderMarker, \timestamp].\epochPublicKey$ and computes $\reputation_\senderMarker \gets \reputation(\DB[\identity_\senderMarker, \timestamp].\score)$.

Next, \accountabilityServer cleans up $\DB[\identity_\senderMarker, \timestamp].\vks$ by removing any key-value pairs where the value (a timestamp) is less than~$\timestamp$.
It checks whether $\commitment_\senderMarker \in \DB[\identity_\senderMarker, \timestamp].\vks.\keys$.
If not, and $|\DB[\identity_\senderMarker, \timestamp].\vks.\keys| \geq \vkLimit$, then \sender cannot use a new verification key at this point and \accountabilityServer returns $\mathsf{Err}$.
Otherwise, it stores $\DB[\identity_\senderMarker, \timestamp].\vks[\commitment_\senderMarker] \gets \timestamp + \mathsf{report\_lock}$ (\autoref{subsec:receiver-setup}).
This mechanism prevents the sender from ever having more than \vkLimit concurrent endorsed channels (with any receiver).

\accountabilityServer samples a ``rerandomization factor'' $s$, uses it to compute a new generator $\generator'$, and derives a tag-ephemeral key \ppEphemeralKey by multiplying the epoch-ephemeral key $\epochPublicKey_\senderMarker$ with~$s$.
This is because we cannot just use \sender's epoch-ephemeral key here; receivers need a corresponding public key to verify that the tags are valid, so using a per epoch key would break \sender's unlinkability (\autoref{subsec:security_privact_obj}).
The tag-ephemeral key could in principle be generated by \sender, but that would require \accountabilityServer to store each tag-ephemeral key, resulting in storage proportional to the number of currently valid tags, which we want to avoid.

\accountabilityServer picks a nonce $\nonce$ and a blinding factor $\blinding$ to prepare a Privacy Pass token request~$\blinded$.
It encrypts $\ciphertext \gets \Encrypt_{\symmetricKey}(\identity_\senderMarker||\nonce||\blinding)$ to be able to link the tag later back to the correct sender, if it is reported.
Both $\nonce$ and $\blinding$ are included to avoid \accountabilityServer having to store this data for each issued tag.

\accountabilityServer signs $\commitment_\senderMarker$, $\commitment_\recipientMarker$, $\timestamp$, $\reputation_\senderMarker$, $\ciphertext$, $\publicParams$, $\blinded$, $\generator'$, and $\ppEphemeralKey$, to produce a signature \signature.
It collects all this data along with the signature to form a tag~$\token$.
\accountabilityServer sends \token to \sender.

Upon receiving and parsing \token, \sender verifies that $\ppEphemeralKey$ is computed from $\generator'$ and multiplies $\blinded$ with its ephemeral secret key $\epochSecretKey_\senderMarker$ to create a \emph{blind sender-token}~\blindSenderToken.
It also creates a zero-knowledge proof \proof that $\blindSenderToken$ was formed correctly.
\sender outputs the full endorsement tag $(\opening_\senderMarker, \opening_\recipientMarker, \verificationKey_\senderMarker, \token, \proof, \blindSenderToken)$.
The full protocol description is given in \autoref{proc:sender_token_tag_gen}. 

\subparagraph{Zero-knowledge proof}
Our protocols use the algorithms $\NIZKDLEQ$ and $\Verify\NIZKDLEQ$ to prove and verify the equality of discrete logarithms in zero knowledge.
We show these algorithms in \autoref{proc:nizk_verif}.
For full details, see~\cite{davidson2018privacy}.

\subparagraph{Sender-token}
To jump ahead a bit, the blind sender-token can later be unblinded by \accountabilityServer using~$\blinding$ (that only it has access to) to create a valid (unblinded) \textit{sender-token}.
The sender-token is a Privacy Pass token that \accountabilityServer can at the end of the epoch show to \sender as evidence of a report.

We note that \sender can verify the validity of the sender-token using just its epoch-ephemeral secret key $\epochSecretKey_\senderMarker$. It does not need to know the rerandomization factor~$s$ that \accountabilityServer used to create the tag-ephemeral key.

\newcommand{\mycolspace}{-1.0cm}
\begin{figure}[ht!]
\procedureblock[colspace=\mycolspace,colsep=0ex,codesize=\scriptsize,bodylinesep=1ex]{\small\senderReportTokenGen}{
    \sender( \publicParams, \epochSecretKey_\senderMarker, \verificationKey_\senderMarker, \handle_\recipientMarker) \< \< \accountabilityServer(\publicParams, (\symmetricKey, \secretKey), \DB, \identity_{\senderMarker}) \< \\ 
    \pcln (q, \generator) \gets \publicParams \< \< \< \\
    \pcln \commitment_\senderMarker, \opening_\senderMarker \gets \Commit(\verificationKey_\senderMarker ) \< \< \< \\
    \pcln \commitment_\recipientMarker, \opening_\recipientMarker \gets \Commit(\handle_\recipientMarker) \< \< \< \\[0.2cm]
    \< \hspace{-1.5cm}\xrightarrow{\hspace{0.5cm}\mbox{\scriptsize$\commitment_\senderMarker, \commitment_\recipientMarker$}\hspace{0.5cm}} \\
    \< \< \pcln \timestamp \gets \timestamp_\text{now} \< \\
    \<\< \pcln \epochPublicKey_\senderMarker \gets \DB[\identity_\senderMarker, \timestamp].\epochPublicKey \< \\
    \<\< \pcln \reputation_\senderMarker \gets \reputation(\DB[\identity_\senderMarker, \timestamp].\score) \< \\
    \<\< \pcln \textbf{\textsf{foreach }} (\commitment_\senderMarker', \timestamp') \in \DB[\identity_\senderMarker, \timestamp].\vks,\\ \<\<\quad\qquad \IfThen{\timestamp' < \timestamp}{\text{delete } (\commitment_\senderMarker', \timestamp')} \\
    \<\< \pcln \IfNewlineThen{ \commitment_\senderMarker \notin \DB[\identity_\senderMarker, \timestamp].\vks.\keys \\ \<\< \qquad\textbf{\textsf{and }} |\DB[\identity_\senderMarker, \timestamp].\vks.\keys| \geq \vkLimit }{\<\< \quad\qquad \return{\Err}}\\
    \<\< \pcln \DB[\identity_\senderMarker, \timestamp].\vks[\commitment_\senderMarker] \\ \<\< \qquad\qquad \gets \timestamp + \mathsf{report\_lock} \\
    \<\< \pcln (q, \generator) \gets \publicParams \< \\
    \<\< \pcln s \gets_\$ \scalarField \\
    \<\< \pcln \generator' \gets s \cdot \generator; \ppEphemeralKey \gets s \cdot \epochPublicKey_s \\
    \< \< \pcln \nonce \gets_\$ \{0, 1\}^*;~ \blinding \gets_\$ \scalarField \< \\
    \< \< \pcln \blinded \gets \blinding \cdot \Hash(\nonce) \< \\
    \< \< \pcln \ciphertext \gets \Encrypt_\symmetricKey(\identity_\senderMarker || \nonce || \blinding) \< \\
    \< \< \pcln \signature \gets \Sign_{\secretKey}(\commitment_\senderMarker || \commitment_\recipientMarker|| \\ \<\< \qquad\qquad \timestamp||\reputation_\senderMarker||\ciphertext||\publicParams||\blinded||\generator'||\ppEphemeralKey) \< \\
    \< \< \pcln \token \gets (\commitment_\senderMarker, \commitment_\recipientMarker, \timestamp, \reputation_\senderMarker,\\ \<\< \qquad\qquad \ciphertext,\publicParams, \blinded, \generator', \ppEphemeralKey, \signature) \< \\
    \< \hspace{-1.5cm}\xleftarrow{\hspace{1.0cm}\mbox{\scriptsize\token}\hspace{1.0cm}} \\
    \pcln \token \to (\commitment_\senderMarker', \commitment_\recipientMarker',\\ \qquad\qquad \timestamp, \reputation_\senderMarker, \ciphertext, \publicParams, \blinded, \generator', \ppEphemeralKey, \signature) \\
    \pcln \IfThen{ \timestamp_\text{now} \not\simeq \timestamp}{\return{\Err}} \\
    \pcln \IfNewlineThen{ \commitment_\senderMarker \neq \commitment_\senderMarker' \textbf{\textsf{ or }}\commitment_\recipientMarker \neq \commitment_\recipientMarker'}{\quad \qquad \return{\Err}} \< \< \< \\ 
    \pcln \mathsf{Verify}_\publicKey(\signature, \token) \< \< \< \\ 
    \pcln \IfThen{\ppEphemeralKey \neq \epochSecretKey_\senderMarker \cdot \generator'}{\return{\Err}}  \<\< \\
    \pcln \blindSenderToken \gets \epochSecretKey_\senderMarker \cdot \blinded \< \< \\
    \pcln \proof \gets \NIZKDLEQ(q, \generator', \ppEphemeralKey, \blinded, \blindSenderToken, \epochSecretKey_\senderMarker) \< \< \< \\
    \pcln \return{(\opening_\senderMarker, \opening_\recipientMarker, \verificationKey_\senderMarker, \token, \proof, \blindSenderToken)} \< \< \<}
\caption{Endorsement tag and blind sender-token issuance protocol.}
\label{proc:sender_token_tag_gen}
\end{figure}

\paragraph{TagReceive}
\sender sends $(\opening_\senderMarker, \opening_\recipientMarker, \verificationKey_\senderMarker, \token, \proof, \blindSenderToken)$ to \recipient, who parses \token, checks that it is no older than $\mathsf{val\_period}$ (recall \autoref{subsec:delayed_reporting}), checks $\commitment_\senderMarker$ and $\commitment_\recipientMarker$ against $(\opening_\senderMarker, \verificationKey_\senderMarker)$ and $(\opening_\recipientMarker, \handle_\recipientMarker)$, respectively, and verifies the signature against \accountabilityServer's~$\publicKey$.
It verifies the zero-knowledge proof \proof against~\blindSenderToken.
If any of these checks fail, it returns $\mathsf{Err}$ as a sign that this was not a valid endorsement tag.

The receiver maintains its dictionary \TagsDict of endorsed channels, as we explained in \autoref{subsec:receiver-setup}.
If $\verificationKey_\senderMarker \notin \TagsDict.\keys$, it writes $\TagsDict[\verificationKey_\senderMarker] \gets (\timestamp_\text{now}, \EmptyQueue)$.
It pushes $(\token, \proof, \blindSenderToken)$ to the queue $\TagsDict[\verificationKey_\senderMarker].\tags$.
The full protocol description is given in \autoref{proc:sender_token_tag_verif}.

\renewcommand{\mycolspace}{-1.0cm}
\begin{figure}[ht!]
\procedureblock[colspace=\mycolspace,colsep=0ex,codesize=\scriptsize,bodylinesep=1ex]{\small\senderReportTokenVerify}
{
    \sender(\opening_\senderMarker, \opening_\recipientMarker, \verificationKey_\senderMarker, \token, \proof, \blindSenderToken) \<\< \recipient(\publicParams, \publicKey, \handle_\recipientMarker, \TagsDict) \< \\[0.2cm]
    \< \hspace{-0.5cm}\xrightarrow{\hspace{0.2cm}\mbox{\scriptsize\ensuremath{\opening_\senderMarker, \opening_\recipientMarker, \verificationKey_\senderMarker, \token, \proof, \blindSenderToken}}\hspace{0.5cm}} \\
    \<\< \pcln (q, \generator) \gets \publicParams \< \\
    \< \< \pcln \token \to (\commitment_\senderMarker, \commitment_\recipientMarker,\\ \<\< \qquad\qquad \timestamp, \reputation_\senderMarker, \ciphertext, \publicParams, \blinded, \generator', \ppEphemeralKey, \signature) \< \\
    \< \< \pcln \IfThen{ \timestamp_\text{now} > \timestamp + \mathsf{val\_period}}{\return{\Err}} \\
    \< \< \pcln \Open(\commitment_\senderMarker, \opening_\senderMarker, \verificationKey_\senderMarker) \< \\
    \< \< \pcln \Open(\commitment_\recipientMarker, \opening_\recipientMarker, \handle_\recipientMarker) \< \\
    \< \< \pcln \Verify_\publicKey(\signature, \commitment_\senderMarker||\commitment_\recipientMarker||\\ \<\<\qquad\qquad \timestamp||\reputation_\senderMarker||\ciphertext||\publicParams ||\blinded||\generator'||\ppEphemeralKey)\< \\
    \< \< \pcln  \Verify\NIZKDLEQ([q, \generator', \ppEphemeralKey, \blinded, \blindSenderToken], \proof) \<  \\
     \<\< \pcln \IfNewlineThen{\verificationKey_\senderMarker \notin \TagsDict.\keys}{ \<\< \qquad\quad  \TagsDict[\verificationKey_\senderMarker].\timestamp_\text{rep} \gets \timestamp_\text{now} } \\
    \<\< \qquad \quad \TagsDict[\verificationKey_\senderMarker].\tags \gets \EmptyQueue
     \\
    \< \< \pcln \text{push } (\token, \proof, \blindSenderToken) \text{ to } \TagsDict[\verificationKey_\senderMarker].\tags \\
}
\caption{Endorsement tag and blinded sender-token verification protocol.}
\label{proc:sender_token_tag_verif}
\end{figure}

\paragraph{TagReport}
If \recipient finds the sender communicating on an endorsed channel $(\verificationKey_\senderMarker, \handle_\recipientMarker)$ to act inappropriately, it reads $(\timestamp_\text{rep}, \tags) \gets \TagsDict[\verificationKey_\senderMarker]$.
If this fails, or $\timestamp_\text{rep}$ is still in the future, it returns~\Err.
Otherwise, it reads (but does not pop) the endorsement tag $ (\token, \proof, \blindSenderToken) \gets \tags.\mathsf{head}$.
It sends this tuple to~\accountabilityServer.

\accountabilityServer parses \token, checks that it has not expired, verifies the signature against \accountabilityServer's $\publicKey$, and verifies \proof against~\blindSenderToken.
If any of these checks fail, it returns $\mathsf{Err}$.
Next, \accountabilityServer decrypts \ciphertext to find the sender's account $\identity_\senderMarker$, $\nonce$, and~$\blinding$.
It unblinds \blindSenderToken by multiplying it with $\blinding^{-1} \bmod{q}$ to obtain a sender-token~$\signature_\senderMarker$.
Finally, it stores $(\nonce, \signature_\senderMarker)$ in the set $\DB[\identity_\senderMarker,\timestamp].\tokens$ for later use as a proof of this report.
The full details of the protocol are in \autoref{proc:report_protocol}.

\renewcommand{\mycolspace}{-1.0cm}
\begin{figure}[ht!]
\procedureblock[colspace=\mycolspace,colsep=0ex,codesize=\scriptsize,bodylinesep=1ex]{\small$\tokenTagReport$}{
    \recipient(\TagsDict, \verificationKey_\senderMarker) \< \< \accountabilityServer(\publicParams, \symmetricKey, \publicKey, \DB) \< \\[0.2cm]
    \pcln \IfNewlineThen{\verificationKey_\senderMarker \notin \TagsDict.\keys}{\qquad\quad\return{\Err}}  \\
    \pcln (\timestamp_\text{rep}, \tags) \gets \TagsDict[\verificationKey_\senderMarker]  \\
    \pcln \IfThen{ \timestamp_\text{now} < \timestamp_\text{rep} }{ \return{\Err} }\\
    \pcln \TagsDict[\verificationKey_\senderMarker].\timestamp_\text{rep} \\
    \quad\qquad \gets \timestamp_\text{now} + \mathsf{report\_lock} \\
    \pcln (\token, \proof, \blindSenderToken) \\
    \quad\qquad \gets \TagsDict[\verificationKey_\senderMarker].\tags.\mathsf{head} \\
    \< \hspace{0cm} \xrightarrow{\hspace{0.7cm}\mbox{\scriptsize\token, \proof, \blindSenderToken}\hspace{0.7cm}} \\
    \<\< \pcln (q, \generator) \gets \publicParams \< \\
    \< \< \pcln \token \to (\commitment_\senderMarker, \commitment_\recipientMarker, \\ \<\< \qquad\qquad \timestamp, \reputation_\senderMarker, \ciphertext,\publicParams ,\blinded, \generator', \ppEphemeralKey, \signature) \< \\
    \< \< \pcln \timestamp_\text{exp} \gets \timestamp + \expTime \cdot \mathsf{epoch\_dur} \\
    \< \< \pcln \IfThen{ \timestamp_\text{now} > \timestamp_\text{exp} }{\return{\Err}} \\
    \< \< \pcln  \Verify_\publicKey(\signature, \commitment_\senderMarker||\commitment_\recipientMarker|| \\ \<\<\qquad\qquad \timestamp||\reputation_\senderMarker||\ciphertext||\publicParams ||\blinded|| \generator' || \ppEphemeralKey) \< \\
   \< \< \pcln  \Verify\NIZKDLEQ(\\
   \< \< \quad\qquad [q, \generator', \ppEphemeralKey, \blinded, \blindSenderToken], \proof) \< \\
   \< \< \pcln  \identity_\senderMarker, \nonce, \blinding \gets \Decrypt_\symmetricKey(\ciphertext) \< \\
   \< \< \pcln \signature_\senderMarker \gets \frac{1}{\blinding}\cdot \blindSenderToken \< \\
   \< \< \pcln   \text{add } (\nonce, \signature_\senderMarker) \text{ to }  \DB[\identity_\senderMarker,\timestamp].\tokens \< \\
}
\caption{Protocol for reporting an endorsement tag and a blind sender-token.}
\label{proc:report_protocol}
\end{figure}

\paragraph{End of epoch}
At the end of an epoch $i$, \accountabilityServer needs to update each sender's score according to~$\reputationFunc$.
For each account identity $\identity_\senderMarker$, \accountabilityServer reads  $\score_i \gets \DB[\identity_\senderMarker, i].\score$, $\vks_i \gets \DB[\identity_\senderMarker, i].\vks$, and $\tokens \gets \DB[\identity_\senderMarker, i-\expTime].\tokens$.
It sets $\reportCount \gets |\tokens|$.
It computes a new score $\score_{i+1} \gets \reputationFunc(\score_i, \reportCount)$. 
It collects each key-value pair $(\verificationKey_\senderMarker, \timestamp) \in \vks_i$ with $\timestamp > \timestamp_\text{now}$ into a new dictionary~$\vks_{i+1}$. 
It writes a new entry $\DB[\identity_\senderMarker, i + 1] \gets (\bot, \score_{i+1}, \vks_{i+1}, \emptyset)$.

If a sender with account $\identity_\senderMarker$ at a later point requests a proof for its score change from epoch $i$ to $i+1$, \accountabilityServer gives it the set~$\tokens$.

For each $(\nonce, \signature_\senderMarker)\in \tokens$, the sender checks that $\nonce$ is unique and that $\signature_\senderMarker = \epochSecretKey_\senderMarker \cdot \Hash(\nonce)$.
If these checks pass, the sender accepts the proof if $\score_{i+1} =  \reputationFunc(\score_i, |\tokens|)$
This interaction with the sender can be postponed until the next time the sender requests new tags.

\section{Reporter Privacy} \label{sec:anonymous_reporting}
The construction in \autoref{sec:protocol} satisfies all of our security and privacy objectives from \autoref{subsec:security_privact_obj} except reporter privacy.
In this section, we discuss how to extend the construction to achieve reporter privacy.
For simplicity, we assume that each receiver uses only a single address~$\handle_\recipientMarker$ for endorsed communication.
In \autoref{subsec:extended_privacy_discussion} we explain how to generalize this to receivers using multiple addresses.

Our reporter privacy objective can be stated as follows:
\begin{center}
\fbox{
\begin{minipage}{0.9\columnwidth}
A sender, identified by $\identity_\senderMarker$, should not be able to learn whether any of its endorsement tags sent to a specific receiver have been reported.
\end{minipage}
}
\end{center}
In some cases, a weaker notion of \textit{report privacy} (as opposed to report\textit{er} privacy) might suffice, where the sender cannot learn whether a specific tag was reported.
This has the practical problem that the receiver's behavior likely correlates across time and eventually provides a statistical signal to the sender that the receiver is, perhaps systematically, reporting it.
If the receiver uses only a single address for endorsed communication, report privacy becomes equivalent to reporter privacy when~$\vkLimit=1$.
We discuss report privacy further in \autoref{subsec:extended_privacy_discussion}.

\subsection{Defining Reporter Privacy}\label{subsec:def-of-report-privacy}
We define reporter privacy first per epoch and then use differential privacy (DP) composition~\cite{dwork2006differential,dwork2014algorithmic,gopi2021numerical} to extend the definition to a longer stretch of time.
For a single epoch, let $\mathcal{M}$ denote a list of channels $(\verificationKey_\senderMarker, \handle_\recipientMarker)$ for which the sender requested, and possibly sent out, endorsement tags during that epoch.
At the end of the epoch the sender is shown a count for the number of reports it received.\footnote{More precisely, with delayed reporting (\autoref{subsec:delayed_reporting}) it is shown the count for reported tags expiring in the presently ending epoch.}
We denote by $\mathcal{D} \subseteq \mathcal{M}$ a subset of tags that were successfully reported so that they appear in the count.

We call two such subsets adjacent if they differ only in tags sent to a single receiver.
Since \accountabilityServer limits the number of addresses the sender can use per epoch and the receiver refrains from reporting tags from the same $\verificationKey_\senderMarker$ more than once within an epoch, two adjacent subsets can differ in size by up to~\vkLimit.
Hence, we need to use DP with sensitivity~$\vkLimit$.
For a subset $\mathcal{D}$, we then define a noisy count as $\mathfrak{R}(\mathcal{D}) := |\mathcal{D}| + \noise$, where $\noise$ is sampled from an appropriate noise distribution~\noiseDistr.
Our per-epoch reporter privacy guarantee states that $\mathfrak{R}$ is $(\epsilon, \delta)$-DP, \textit{i.e.}, for any two adjacent subsets $\mathcal{D}_0, \mathcal{D}_1 \subseteq \mathcal{M}$ and any subset $\mathcal{S}$ of possible noisy report counts,
\[ \Pr[\mathfrak{R}(\mathcal{D}_0) \in \mathcal{S}] \leq e^{\epsilon} \Pr[\mathfrak{R}(\mathcal{D}_1) \in \mathcal{S}] + \delta\,. \]

The above definition is for a per-epoch privacy guarantee.
We extend to a longer period through DP composition using the techniques in~\cite{prv_accountant}.

\subsection{Implementing Reporter Privacy}\label{sec:report_privacy_with_trusted_as}
Next, we explain how exactly to implement reporter privacy.
To be clear, we assume malicious senders and a malicious \accountabilityServer do not collude against a receiver to break reporter privacy.
Moreover, we want to ensure we continue protecting against a malicious \accountabilityServer trying to break a sender's score integrity.

\paragraph{Negative noise}
\accountabilityServer samples a noise $\noise \gets \noiseDistr$ and adds \noise to the true report count $\reportCount$, presenting $\reportCount + \noise$ sender-tokens to the sender as a proof of received reports.
Since \accountabilityServer cannot create new reports by itself, we limit to using only negative noise, \textit{i.e.}, the \accountabilityServer omits some of the sender-tokens when presenting its evidence.
Concretely, \noiseDistr is a Gaussian distribution that is left-shifted by $\noiseMean < -1$, truncated to $(-\infty, -1]$, and rounded to the nearest integer.\footnote{We show in \autoref{appendix:scale_noise} that when the sensitivity changes from $1$ to $\vkLimit$, the noise distribution changes from $\noiseDistr \left( \noiseMean, \sigma^2, - 1/2\right)$ to $\noiseDistr' \left( -1/2 + \vkLimit \left( 1/2 + \noiseMean \right), (\vkLimit \cdot \sigma)^2, -1/2\right)$. Our notation for $\noiseDistr$ indicates the mean $\noiseMean$, variance $\sigma^2$, and truncation point $-1/2$ due to rounding to the nearest integer.}

If \accountabilityServer is malicious, it can choose \noise to be smaller (more negative) than it would have honestly sampled, but this only benefits the sender.
If it samples \noise larger (closer to $-1$) than it would have honestly sampled, it reduces the sender's benefit, but it cannot change the protection guaranteed by the tolerance level~\bonus.
Thus, the impact of a malicious \accountabilityServer is limited.

\paragraph{End of epoch}
Ending an epoch works mostly as in \autoref{subsec:protocol_interface}.
The only difference is that, for each sender, \accountabilityServer samples a noise from \noiseDistr and computes a new score using the noisy report count.
When the sender asks for a proof of its report count, \accountabilityServer shows it $|\tokens| + \noise$ sender-tokens from~$\mathsf{tokens}$.
Since $\noise \leq -1$, \accountabilityServer simply omits some subset of the received tokens.
If $|\tokens| +\noise \leq 0$, it shows no tokens.

The sender validates the sender-tokens as in \autoref{subsec:protocol_interface}.
\accountabilityServer can catch up an inactive sender using $\reportCount = 0$ and $\noise = \noiseMean$ for each of the epochs the sender did not interact with the system.

\paragraph{Examples}
We use \cite{prv_accountant} for numerical estimations of DP composition.
For $\vkLimit = 1$, to reach $(\epsilon=1, \delta=2^{-16})$ we need $\noiseMean = -17$ and $\sigma = 3.7$, and to reach $(\epsilon=4, \delta=2^{-16})$ we need $\noiseMean = -8$ and $\sigma = 1.1$.
For example, using $\bonus = 1$ in the first case and $\bonus = 10$ in the second case both result in roughly 18 reports being ignored on average, but in the first case only 2 reports are certainly ignored, whereas in the second case 11 reports are certainly ignored.

We present more experiments in \autoref{tab:noise-parameters}, varying the number of compositions and $\vkLimit$, but keeping $(\epsilon=4, \delta=2^{-16})$ fixed.

\begin{table}[h!]
    \footnotesize
    \centering
    \begin{tabular}{|c|c|c|c|}
        \hline
        $\vkLimit$ & Compositions & $\noiseMean$ & $\sigma$  \\
        \hline\hline
        1 & 1 & -8 & 1.1 \\ \hline
        1 & 10 & -18 & 3.5 \\ \hline
        1 & 20 & -30 & 5 \\ \hline
        1 & 40 & -40 & 7 \\ \hline
        1 & 100 & -50 & 11 \\ \hline
        3 & 1 & -23 & 3.3 \\ \hline
        3 & 10 & -40 & 10 \\ \hline
        5 & 1 & -38 & 5.5 \\
        \hline
    \end{tabular}
    \caption{DP parameter settings for different numbers of compositions. In all cases we get $(\epsilon=4, \delta=2^{-16})$.}
    \label{tab:noise-parameters}
\end{table}

\section{Optimality}\label{sec:optimality}
We describe now how a rational sender (\autoref{def:senders-rationality}) using \oursystem will behave and show how its behavior is controlled by the tolerance level \bonus in the definition of \reputationFunc (\autoref{def:score_function}).
We assume all receivers to be rational (\autoref{def:receivers-rationality}).

To simplify the analysis, we assume no delayed reporting, \textit{i.e.}, $\expTime = 0$.
This means that all reports are received immediately by \accountabilityServer and have an effect at the next epoch change.
In \autoref{subsec:optimality-further-results}, we show how to incorporate delayed reporting.

We assume all channels are endorsed\footnote{This limitation is necessary without placing some cost or limit to the number of non-endorsed messages that can be sent within an epoch.} and that all scripts are executed instantaneously.
We assume that receivers make reporting decisions and submit their reports instantaneously.
We make a worst-case assumption, providing advantage to an adversarial sender, that the sender learns immediately if it was reported.
We assume \messageSet satisfies the properties in \autoref{sec:props-of-script-space}.

\subsection{Memoryless Game}\label{subsec:memoryless-game}
We consider a game for a rational sender, where in each move the sender has two options: (1) it can set up a new endorsed channel and use it to execute a script within the current epoch, or (2) it can wait for the next epoch.
In each epoch, the sender executes one script per each endorsed channel.
At the end of epoch $i$ the sender's score is computed from its current score \score, its number of reports \reportCount, and the noise $\noise_i$ as $\reputationFunc(\score, \reportCount + \noise_i)$.

We assume that receivers are memoryless and probabilistic so that their behavior is entirely captured by the random variables $\rewardLabel$, $\probReward$, and $\probReport$ (recall \autoref{subsec:conversations-and-scripts}).
We assume the number of epochs is limited, as the senders or \accountabilityServer do not exist forever.
We assume \accountabilityServer's random coins are fixed and denote by $\noise_i$ the noise it samples in epoch~$i$.
We assume that there is a limit \remainingReports to the total number of reports each sender is allowed to get during the entire game (including DP noise), after which it can get no more tags or rewards.
Thus, a rational sender should use its entire report budget \remainingReports in maximizing its expected total reward.

\subsection{Sender's Strategy}
The sender's state is a tuple $(\score, \reportCount, \epoch, \remainingReports)$, where \score is its current score, \reportCount is the number of reports it has received so far in the current epoch, \epoch is the number of remaining epochs in the game excluding the current epoch, and \remainingReports is as explained above.
We denote by $\optFunc_{\score, \epoch, \remainingReports}^{\reportCount}$ the sender's maximal expected total reward from now on, given its state $(\score, \reportCount, \epoch, \remainingReports)$.

With this notation we can finally formalize a rational sender's behavior to ``maximize its expected total reward'' in \autoref{def:senders-rationality}.
\begin{definition}[Optimal sender's strategy]\label{def:optimal_senders_strategy}
A sender's strategy is defined by a function mapping $(\score, \epoch, \remainingReports, \reportCount)$ to the next move, so that the move is always to wait when $\remainingReports =0$.
We denote $\optFunc(\score, \epoch, \remainingReports) := \optFunc_{\score, \epoch, \remainingReports}^0$ and say that a sender's strategy is optimal if it reaches $\optFunc(\score, \epoch, \remainingReports)$.
\end{definition}
A particularly interesting class of sender's strategies are what we call \textit{normalized} strategies.
While the definition will seem odd at this time, its importance becomes obvious with \autoref{th:opt_strategy_1} and \autoref{th:opt_strategy_2}.
\begin{definition}[Normalized sender's strategy]\label{def:normalized_strategy}
A normalized sender's strategy is a sequence of non-negative integers $(X_\epoch, \ldots, X_0)$.
If $X_i = 0$, the sender executes no scripts in epoch $i$ and instead waits until the next epoch.
If $X_i > 0$, the sender executes \textit{different} scripts $\content$, each maximizing $\widetilde{\rewardLabel}(\content, \reputation_\senderMarker)$, until $X_i$ reports are received, then waits until the next epoch.
\label{def:sender_strategy}
\end{definition}
Finally we can state our two main optimality theorems.
We prove them in \autoref{subsec:proof-of-th1} and \autoref{subsec:proof-of-th2}, respectively.
\begin{theorem}\label{th:opt_strategy_1}
    Given our assumptions, for any $(\score, \epoch, \remainingReports)$, there exists an optimal sender's strategy.
    Moreover, any optimal strategy is normalized.
\end{theorem}
\begin{theorem}
    Given a score function $\reputationFunc_\bonus^{\maxScore, \reportScale}$ and a reputation function~$\reputation$, for any $(\score, \epoch, \remainingReports)$, and any normalized strategy $(X_\epoch, \ldots, X_0)$, there exists always at least as good of a normalized strategy, where $0 \leq X_i + \noise_i \leq \bonus$ for all $i \geq 1$. Here $\noise_i$ denotes the noise in epoch~$i$.

    There exists an optimal normalized strategy $(X_\epoch^\text{opt}, \ldots, X_0^\text{opt})$, where $0 \leq X_i^\text{opt} + \noise_i \leq \bonus$ for all~$i \geq 1$.
\label{th:opt_strategy_2}
\end{theorem}

\paragraph{Discussion}
\autoref{th:opt_strategy_2} lets \accountabilityServer reduce $\bonus$ to incentivize rational senders to try to receive a smaller number of reports until they stop using \oursystem.
The (negative) noise values limit its ability to do this efficiently; this is the privacy-utility trade-off common in~DP.
\accountabilityServer can optionally choose to bound the noise distribution from below to arrive at a uniform bound for the~$X_i$.
The downside is that this either weakens the DP guarantee or provides (in practice) a very loose bound on the~$X_i$.
Without such uniform bounding, the expected upper bound for each $X_i$ is $\bonus - \mathbb{E}[\noiseDistr] \approx \bonus - \noiseMean$.

Note that \autoref{th:opt_strategy_2} does not guarantee that there would not exist an optimal strategy outside this bounded region for the~$X_i$.
Rather, its power is that it gives the sender a tight search space for an optimal strategy, so that the smaller \bonus is (smallest value is 1) the smaller this search space is.
The flip-side is that if the sender's strategy falls outside of this region it becomes unclear how far it deviates from an optimal strategy.

In \autoref{subsec:optimality-further-results} we analyze how close the sender can get to its optimal reward by merely ensuring its strategy falls within the bounds in \autoref{th:opt_strategy_2}, even if it cannot follow the exact optimal strategy.

\section{Implementation}\label{sec:implementation}
We implemented the protocols \autoref{proc:epoch_key_reg}, \autoref{proc:sender_token_tag_gen}, \autoref{proc:sender_token_tag_verif}, and \autoref{proc:report_protocol} in Rust and ran them on a \texttt{Standard\_E16ads\_v5} Azure VM (AMD EPYC 7763 @ 2.445~GHz). Our benchmarks ran on a single thread.

We used HMAC-SHA256 for commitments and Curve25519 for elliptic curve operations (from \texttt{curve25519‑dalek}), including \NIZKDLEQ and $\Verify\NIZKDLEQ$ (\autoref{appendix:crypto_prelim}).
For digital signatures, we used $\mathsf{EdDSA}$ (from \texttt{ed25519‑dalek}).
For symmetric key encryption we used AES-CBC.
Our code and benchmarks are available (MIT license) at \url{https://GitHub.com/Microsoft/Sandi}.

The serialized data from $\accountabilityServer$ to the sender ($\token$ in \autoref{proc:sender_token_tag_gen}) was $304\,$B and the full endorsement tag, $(\opening_\senderMarker, \opening_\recipientMarker, \verificationKey_\senderMarker, \token, \proof, \blindSenderToken)$, was $508\,$B. Memory storage requirements are shown in \autoref{tab:storage-sizes}.
The running times are shown in \autoref{tab:function-times} for the different protocols and different parties.

\begin{table}[htbp]
\footnotesize
\centering
\begin{tabular}{|c|c|c|}
\hline
\textbf{From} & \textbf{To} & \textbf{Size} \\
\hline
\textbf{\accountabilityServer} & \sender & $304\,$B per requested tag \\
\hline
\textbf{\sender} & \recipient & $508\,$B per requested tag \\
\hline
\end{tabular}
\caption{Communication sizes}
\label{tab:communication-sizes}
\end{table}

\begin{table}[htbp]
\footnotesize
\centering
\begin{tabular}{|c|c|}
\hline
\textbf{Measurements on} & \textbf{Memory requirement} \\
\hline
\textbf{\accountabilityServer} & $160\,$B per sender \\
\hline
\textbf{\sender} & $248\,$B per channel \\
\hline
\end{tabular}
\caption{Storage sizes}
\label{tab:storage-sizes}
\end{table}

\begin{table}[htbp]
\footnotesize
\centering
\begin{tabular}{|c|c|c|}
\hline
\textbf{Protocol} & \textbf{Party} & \textbf{Total Time} \\
\hline
\multirow{2}{*}{\senderReportTokenGen} & \sender & $171\,\mu$s \\
\cline{2-3}
 & \textbf{\accountabilityServer} & $158\,\mu$s ($97\,\mu$s in \NIZKDLEQ) \\
\hline
\senderReportTokenVerify & \recipient & $234\,\mu$s ($162\,\mu$s in $\Verify\NIZKDLEQ$) \\
\hline
\tokenTagReport & \accountabilityServer & $279\,\mu$s ($162\,\mu$s in $\Verify\NIZKDLEQ$) \\
\hline
\end{tabular}
\caption{Summary of protocol execution times.}
\label{tab:function-times}
\end{table}

\section{Conclusions}
We have presented a novel reputation system, \oursystem, that is designed to create trust through accountability for one-to-one communication.
It addresses the shortcomings of traditional reputation systems with its strong security and privacy properties and by preventing the emergence of undesirable optimal strategies.

There are many avenues for future work to be explored, including exploring other possible uses for a reputation system like \oursystem, extending \oursystem to support a notion of correctness for receivers, and enabling senders to publicly prove if \accountabilityServer is misbehaving without requiring them to maintain a long-term cryptographic state.

\newcommand{\ackstext}{
We thank the numerous people at Microsoft that were involved in discussions around this work.
Specifically, we would like to thank Melissa Chase for long and invaluable conversations, Shrey Jain and Robert Sim for close collaboration, and Ofer Dekel for guiding us in the right direction.
We would like to thank Hamed Khanpour, Graham Reeve, and Harsha Nori for many helpful discussions.
}

\section*{Acknowledgments} \ackstext

\bibliographystyle{plain}
\bibliography{bibliography}

\begin{thebibliography}{10}

\bibitem{hashcash}
Adam Back.
\newblock Hashcash -- {A} {D}enial of {S}ervice {C}ounter-{M}easure.
\newblock \url{http://www.hashcash.org/papers/hashcash.pdf}. Accessed on 1/21/2024, 2002.

\bibitem{bader2023reputation}
Lennart Bader, Jan Pennekamp, Emildeon Thevaraj, Maria Spiß, Salil~S. Kanhere, and Klaus Wehrle.
\newblock Reputation systems for supply chains: The challenge of achieving privacy preservation, 2023.

\bibitem{bag2018privacy}
Samiran Bag, Muhammad~Ajmal Azad, and Feng Hao.
\newblock A privacy-aware decentralized and personalized reputation system.
\newblock {\em Computers \& Security}, 77:514--530, 2018.

\bibitem{bemmann2018fully}
Kai Bemmann, Johannes Bl{\"o}mer, Jan Bobolz, Henrik Br{\"o}cher, Denis Diemert, Fabian Eidens, Lukas Eilers, Jan Haltermann, Jakob Juhnke, Burhan Otour, et~al.
\newblock Fully-featured anonymous credentials with reputation system.
\newblock In {\em Proceedings of the 13th International Conference on Availability, Reliability and Security}, pages 1--10, 2018.

\bibitem{bernstein2013elligator}
Daniel~J Bernstein, Mike Hamburg, Anna Krasnova, and Tanja Lange.
\newblock {Elligator: elliptic-curve points indistinguishable from uniform random strings.}
\newblock In {\em Proceedings of the 2013 ACM SIGSAC conference on Computer \& communications security}, pages 967--980, 2013.

\bibitem{blomer2018practical}
Johannes Bl{\"o}mer, Fabian Eidens, and Jakob Juhnke.
\newblock Practical, anonymous, and publicly linkable universally-composable reputation systems.
\newblock In {\em Topics in Cryptology--CT-RSA 2018: The Cryptographers' Track at the RSA Conference 2018, San Francisco, CA, USA, April 16-20, 2018, Proceedings}, pages 470--490. Springer, 2018.

\bibitem{blomer2015anonymous}
Johannes Bl{\"o}mer, Jakob Juhnke, and Christina Kolb.
\newblock Anonymous and publicly linkable reputation systems.
\newblock In {\em International Conference on Financial Cryptography and Data Security}, pages 478--488. Springer, 2015.

\bibitem{pow_cant_can_does_work}
L.~Jean Camp and Debin Liu.
\newblock Proof of {W}ork ({C}annot, {C}an, {D}oes {C}urrently) {W}ork.
\newblock \url{https://ssrn.com/abstract=2118235}. Accessed on 10/15/2023, 2012.

\bibitem{chirita2005mailrank}
Paul-Alexandru Chirita, J{\"o}rg Diederich, and Wolfgang Nejdl.
\newblock Mailrank: using ranking for spam detection.
\newblock In {\em Proceedings of the 14th ACM international conference on Information and knowledge management}, pages 373--380, 2005.

\bibitem{clauss2013k}
Sebastian Clau{\ss}, Stefan Schiffner, and Florian Kerschbaum.
\newblock K-anonymous reputation.
\newblock In {\em Proceedings of the 8th ACM SIGSAC symposium on Information, computer and communications security}, pages 359--368, 2013.

\bibitem{cormack2008email}
Gordon~V Cormack et~al.
\newblock {Email spam filtering: A systematic review.}
\newblock {\em Foundations and Trends{\textregistered} in Information Retrieval}, 1(4):335--455, 2008.

\bibitem{dada2019machine}
Emmanuel~Gbenga Dada, Joseph~Stephen Bassi, Haruna Chiroma, Adebayo~Olusola Adetunmbi, Opeyemi~Emmanuel Ajibuwa, et~al.
\newblock {Machine learning for email spam filtering: review, approaches and open research problems.}
\newblock {\em Heliyon}, 5(6), 2019.

\bibitem{davidson2018privacy}
Alex Davidson, Ian Goldberg, Nick Sullivan, George Tankersley, and Filippo Valsorda.
\newblock Privacy pass: Bypassing internet challenges anonymously.
\newblock {\em Proc. Priv. Enhancing Technol.}, 2018(3):164--180, 2018.

\bibitem{dwork2006differential}
Cynthia Dwork.
\newblock Differential privacy.
\newblock In {\em International colloquium on automata, languages, and programming}, pages 1--12. Springer, 2006.

\bibitem{pricing_via_combatting_junk_mail}
Cynthia Dwork and Moni Naor.
\newblock Pricing via processing or combatting junk mail.
\newblock In {\em Advances in Cryptology --- CRYPTO' 92}, pages 139--147, Berlin, Heidelberg, 1993. Springer Berlin Heidelberg.

\bibitem{dwork2014algorithmic}
Cynthia Dwork, Aaron Roth, et~al.
\newblock {The algorithmic foundations of differential privacy.}
\newblock {\em Foundations and Trends{\textregistered} in Theoretical Computer Science}, 9(3--4):211--407, 2014.

\bibitem{esquivel2010effectiveness}
Holly Esquivel, Aditya Akella, and Tatsuya Mori.
\newblock On the effectiveness of ip reputation for spam filtering.
\newblock In {\em 2010 Second International Conference on COMmunication Systems and NETworks (COMSNETS 2010)}, pages 1--10. IEEE, 2010.

\bibitem{farmer2010building}
Randy Farmer and Bryce Glass.
\newblock {\em Building web reputation systems}.
\newblock "O'Reilly Media, Inc.", 2010.

\bibitem{friedman2001social}
Eric~J Friedman* and Paul Resnick.
\newblock The social cost of cheap pseudonyms.
\newblock {\em Journal of Economics \& Management Strategy}, 10(2):173--199, 2001.

\bibitem{goodrich2011privacy}
Michael~T Goodrich and Florian Kerschbaum.
\newblock Privacy-enhanced reputation-feedback methods to reduce feedback extortion in online auctions.
\newblock In {\em Proceedings of the first ACM conference on Data and application security and privacy}, pages 273--282, 2011.

\bibitem{gopi2021numerical}
Sivakanth Gopi, Yin~Tat Lee, and Lukas Wutschitz.
\newblock Numerical composition of differential privacy.
\newblock {\em Advances in Neural Information Processing Systems}, 34:11631--11642, 2021.

\bibitem{prv_accountant}
Sivakanth Gopi, Yin~Tat Lee, and Lukas Wutschitz.
\newblock Privacy {R}andom {V}ariable {(PRV) {A}ccountant}.
\newblock \url{https://github.com/microsoft/prv_accountant}. Accessed on 10/15/2023, 2022.

\bibitem{grier2010spam}
Chris Grier, Kurt Thomas, Vern Paxson, and Michael Zhang.
\newblock {@ spam: the underground on 140 characters or less}.
\newblock In {\em Proceedings of the 17th ACM conference on Computer and communications security}, pages 27--37, 2010.

\bibitem{gurtler2021sok}
Stan Gurtler and Ian Goldberg.
\newblock {SoK: Privacy-Preserving Reputation Systems.}
\newblock {\em Proc. Priv. Enhancing Technol.}, 2021(1):107--127, 2021.

\bibitem{guzella2009review}
Thiago~S Guzella and Walmir~M Caminhas.
\newblock {A review of machine learning approaches to spam filtering.}
\newblock {\em Expert Systems with Applications}, 36(7):10206--10222, 2009.

\bibitem{money_burning_optimal_mechanism}
Jason~D Hartline and Tim Roughgarden.
\newblock Optimal mechanism design and money burning.
\newblock In {\em Proceedings of the fortieth annual ACM symposium on Theory of computing}, pages 75--84, 2008.

\bibitem{hasan2022privacy}
Omar Hasan, Lionel Brunie, and Elisa Bertino.
\newblock {Privacy-Preserving Reputation Systems based on Blockchain and Other Cryptographic Building Blocks: A Survey.}
\newblock {\em ACM Computing Surveys (CSUR)}, 55(2):1--37, 2022.

\bibitem{hauser2023street}
Christophe Hauser, Shirin Nilizadeh, Yan Shoshitaishvili, Ni~Trieu, Srivatsan Ravi, Christopher Kruegel, and Giovanni Vigna.
\newblock Street rep: A privacy-preserving reputation aggregation system.
\newblock {\em Cryptology ePrint Archive}, 2023.

\bibitem{he2022market}
Sherry He, Brett Hollenbeck, and Davide Proserpio.
\newblock The market for fake reviews.
\newblock {\em Marketing Science}, 41(5):896--921, 2022.

\bibitem{hendrikx2015reputation}
Ferry Hendrikx, Kris Bubendorfer, and Ryan Chard.
\newblock {Reputation Systems: A Survey and Taxonomy.}
\newblock {\em Journal of Parallel and Distributed Computing}, 75:184--197, 2015.

\bibitem{hoffman2009survey}
Kevin Hoffman, David Zage, and Cristina Nita-Rotaru.
\newblock {A Survey of Attack and Defense Techniques for Reputation Systems.}
\newblock {\em ACM Computing Surveys (CSUR)}, 42(1):1--31, 2009.

\bibitem{issa2022hecate}
Rawane Issa, Nicolas Alhaddad, and Mayank Varia.
\newblock {Hecate: Abuse Reporting in Secure Messengers with Sealed Sender.}
\newblock In {\em 31st USENIX Security Symposium (USENIX Security 22)}, pages 2335--2352, 2022.

\bibitem{josang2002beta}
Audun Josang and Roslan Ismail.
\newblock {The Beta Reputation System.}
\newblock In {\em Proceedings of the 15th bled electronic commerce conference}, volume~5, pages 2502--2511. Citeseer, 2002.

\bibitem{josang2007survey}
Audun J{\o}sang, Roslan Ismail, and Colin Boyd.
\newblock {A Survey of Trust and Reputation Systems for Online Service Provision.}
\newblock {\em Decision support systems}, 43(2):618--644, 2007.

\bibitem{josefsson2017edwards}
Simon Josefsson and Ilari Liusvaara.
\newblock {Edwards-Curve Digital Signature Algorithm (EdDSA).}
\newblock RFC 8032, January 2017.

\bibitem{kamara2022outside}
Seny Kamara, Mallory Knodel, Emma Llans{\'o}, Greg Nojeim, Lucy Qin, Dhanaraj Thakur, and Caitlin Vogus.
\newblock Outside looking in: Approaches to content moderation in end-to-end encrypted systems.
\newblock {\em arXiv preprint arXiv:2202.04617}, 2022.

\bibitem{kerschbaum2009verifiable}
Florian Kerschbaum.
\newblock A verifiable, centralized, coercion-free reputation system.
\newblock In {\em Proceedings of the 8th ACM Workshop on Privacy in the Electronic Society}, pages 61--70, 2009.

\bibitem{IETF-SPF}
Scott Kitterman.
\newblock {Sender Policy Framework (SPF) for Authorizing Use of Domains in Email, Version 1}.
\newblock RFC 7208, April 2014.

\bibitem{koutrouli2012taxonomy}
Eleni Koutrouli and Aphrodite Tsalgatidou.
\newblock Taxonomy of attacks and defense mechanisms in p2p reputation systems—lessons for reputation system designers.
\newblock {\em Computer Science Review}, 6(2-3):47--70, 2012.

\bibitem{kozlov2020evaluating}
Fedor Kozlov, Isabella Yuen, Jakub Kowalczyk, Daniel Bernhardt, David Freeman, Paul Pearce, and Ivan Ivanov.
\newblock Evaluating changes to fake account verification systems.
\newblock In {\em 23rd International Symposium on Research in Attacks, Intrusions and Defenses (RAID 2020)}, pages 135--148, 2020.

\bibitem{krebs2014spam}
Brian Krebs.
\newblock {\em Spam nation: The inside story of organized cybercrime-from global epidemic to your front door}.
\newblock Sourcebooks, Inc., 2014.

\bibitem{IETF-DKIM}
Murray Kucherawy, Dave Crocker, and Tony Hansen.
\newblock {DomainKeys Identified Mail (DKIM) Signatures}.
\newblock RFC 6376, September 2011.

\bibitem{IETF-DMARC}
Murray Kucherawy and Elizabeth Zwicky.
\newblock {Domain-based Message Authentication, Reporting, and Conformance (DMARC)}.
\newblock RFC 7489, March 2015.

\bibitem{pow_cant_work}
Ben Laurie and Richard Clayton.
\newblock Proof-of-work proves not to work; version 0.2.
\newblock In {\em Workshop on economics and information, security}, 2004.

\bibitem{liu2006proof}
Debin Liu and L~Jean Camp.
\newblock {Proof of Work can Work.}
\newblock In {\em WEIS}. Citeseer, 2006.

\bibitem{facts2022}
Linsheng Liu, Daniel~S. Roche, Austin Theriault, and Arkady Yerukhimovich.
\newblock {Fighting Fake News in Encrypted Messaging with the Fuzzy Anonymous Complaint Tally System {(FACTS)}.}
\newblock In {\em 29th Annual Network and Distributed System Security Symposium, {NDSS} 2022, San Diego, California, USA, April 24-28, 2022}. The Internet Society, 2022.

\bibitem{marwick2011tweet}
Alice~E Marwick and Danah Boyd.
\newblock {I tweet honestly, I tweet passionately: Twitter users, context collapse, and the imagined audience.}
\newblock {\em New media \& society}, 13(1):114--133, 2011.

\bibitem{mirkovic2008building}
Jelena Mirkovic and Peter Reiher.
\newblock Building accountability into the future internet.
\newblock In {\em 2008 4th Workshop on Secure Network Protocols}, pages 45--51. IEEE, 2008.

\bibitem{movschovitz2013analysis}
Dana Movshovitz-Attias, Yair Movshovitz-Attias, Peter Steenkiste, and Christos Faloutsos.
\newblock Analysis of the reputation system and user contributions on a question answering website: Stackoverflow.
\newblock pages 886--893, 08 2013.

\bibitem{petrlic2014privacy}
Ronald Petrlic, Sascha Lutters, and Christoph Sorge.
\newblock Privacy-preserving reputation management.
\newblock In {\em Proceedings of the 29th Annual ACM Symposium on Applied Computing}, pages 1712--1718, 2014.

\bibitem{meta_message_franking}
Meta Platforms.
\newblock Messenger {S}ecret {C}onversations ({T}echnical {W}hitepaper version 2.0).
\newblock \url{https://about.fb.com/wp-content/uploads/2016/07/messenger-secret-conversations-technical-whitepaper.pdf}. Accessed on 10/15/2023, 2017.

\bibitem{prakash2005fighting}
Vipul~Ved Prakash and Adam O'Donnell.
\newblock Fighting spam with reputation systems: User-submitted spam fingerprints.
\newblock {\em Queue}, 3(9):36--41, 2005.

\bibitem{ramachandran2007filtering}
Anirudh Ramachandran, Nick Feamster, and Santosh Vempala.
\newblock Filtering spam with behavioral blacklisting.
\newblock In {\em Proceedings of the 14th ACM conference on computer and communications security}, pages 342--351, 2007.

\bibitem{resnick2002trust}
Paul Resnick and Richard Zeckhauser.
\newblock Trust among strangers in internet transactions: Empirical analysis of ebay's reputation system.
\newblock In {\em The Economics of the Internet and E-commerce}, pages 127--157. Emerald Group Publishing Limited, 2002.

\bibitem{schaub2016trustless}
Alexander Schaub, R{\'e}mi Bazin, Omar Hasan, and Lionel Brunie.
\newblock A trustless privacy-preserving reputation system.
\newblock In {\em ICT Systems Security and Privacy Protection: 31st IFIP TC 11 International Conference, SEC 2016, Ghent, Belgium, May 30-June 1, 2016, Proceedings 31}, pages 398--411. Springer, 2016.

\bibitem{schiffner2011limits}
Stefan Schiffner, Andreas Pashalidis, and Elmar Tischhauser.
\newblock On the limits of privacy in reputation systems.
\newblock In {\em Proceedings of the 10th annual ACM workshop on Privacy in the electronic society}, pages 33--42, 2011.

\bibitem{SchlegelSpam}
Roman Schlegel and Serge Vaudenay.
\newblock Enforcing email addresses privacy using tokens.
\newblock In {\em Information Security and Cryptology}, pages 91--100. Springer Berlin, 2005.

\bibitem{sekar2023technical}
Shreyas Sekar.
\newblock Technical perspective: A rare glimpse of tracking fake reviews.
\newblock {\em Communications of the ACM}, 66(10):97--97, 2023.

\bibitem{steinbrecher2006design}
Sandra Steinbrecher.
\newblock Design options for privacy-respecting reputation systems within centralised internet communities.
\newblock In {\em IFIP International Information Security Conference}, pages 123--134. Springer, 2006.

\bibitem{tyagi2019asymmetric}
Nirvan Tyagi, Paul Grubbs, Julia Len, Ian Miers, and Thomas Ristenpart.
\newblock {Asymmetric Message Franking: Content Moderation for Metadata-Private End-to-End Encryption.}
\newblock In {\em Advances in Cryptology--CRYPTO 2019: 39th Annual International Cryptology Conference, Santa Barbara, CA, USA, August 18--22, 2019, Proceedings, Part III 39}, pages 222--250. Springer, 2019.

\bibitem{wang2022large}
Chuhan Wang, Kaiwen Shen, Minglei Guo, Yuxuan Zhao, Mingming Zhang, Jianjun Chen, Baojun Liu, Xiaofeng Zheng, Haixin Duan, Yanzhong Lin, et~al.
\newblock A large-scale and longitudinal measurement study of {DKIM} deployment.
\newblock In {\em 31st USENIX Security Symposium (USENIX Security 22)}, pages 1185--1201, 2022.

\bibitem{pow_spam_filter}
Chidi Williams.
\newblock Proof-of-{W}ork {S}pam {F}ilter.
\newblock \url{https://chidiwilliams.com/posts/the-proof-of-work-spam-filter}. Accessed on 1/21/2024, 2021.

\bibitem{xiao2016survey}
Zhifeng Xiao, Nandhakumar Kathiresshan, and Yang Xiao.
\newblock A survey of accountability in computer networks and distributed systems.
\newblock {\em security and communication networks}, 9(4):290--315, 2016.

\bibitem{zhai2016anonrep}
Ennan Zhai, David~Isaac Wolinsky, Ruichuan Chen, Ewa Syta, Chao Teng, and Bryan Ford.
\newblock Anon{R}ep: Towards {T}racking-{R}esistant {A}nonymous {R}eputation.
\newblock In {\em 13th USENIX Symposium on Networked Systems Design and Implementation (NSDI 16)}, pages 583--596, 2016.

\bibitem{zhang2009ipgrouprep}
Hong Zhang, Haixin Duan, Wu~Liu, and Jianping Wu.
\newblock Ipgrouprep: A novel reputation based system for anti-spam.
\newblock In {\em 2009 Symposia and Workshops on Ubiquitous, Autonomic and Trusted Computing}, pages 513--518. IEEE, 2009.

\bibitem{zheleva2008trusting}
Elena Zheleva, Aleksander Kolcz, and Lise Getoor.
\newblock Trusting spam reporters: A reporter-based reputation system for email filtering.
\newblock {\em ACM Transactions on Information Systems (TOIS)}, 27(1):1--27, 2008.

\end{thebibliography}

\inappendixtrue
\newcommand{\appendixsection}{\section}
\newcommand{\appendixsubsection}{\subsection}
\appendices

\appendixsection{Cryptographic Background} \label{appendix:crypto_prelim}

\paragraph{Cryptographic commitment scheme}
A cryptographic commitment scheme consists of two algorithms, $\Commit$, which creates a commitment $\commitment$ and opening $\opening$ for a message $\content$.
Another algorithm $\Open$ can be used to verify this relationship.
The scheme should be \textit{binding}, so that using $\Open$ to produce a different $\content$ is computationally infeasible, and \textit{hiding}, so that $\commitment$ reveals negligible information about~$\content$.
$$\Commit(\content) \to (\commitment, \opening)$$
$$\Open(\commitment, \opening, \content) \to \top / \bot$$
An efficient commitment scheme can be implemented using a cryptographic hash function.

\paragraph{Digital signature scheme}
A digital signature scheme is an asymmetric algorithm, where a public key $\publicKey$ can be used to verify that a message was signed by a party holding a corresponding secret key $\secretKey$.
The signature $\signature$ is generated with a signing algorithm $\Sign$, and verified with a verification algorithm $\Verify$.

$$\AsymmetricKeyGen(1^\securityFactor) \to (\secretKey, \publicKey)$$
$$\Sign_\secretKey (\content) \to \sigma$$
$$\Verify_\publicKey (\sigma, \content) \to \top / \bot$$

A modern standard implementation is the EdDSA algorithm~\cite{josefsson2017edwards}.

\paragraph{Symmetric encryption scheme}
A symmetric encryption scheme consists of two algorithms, $\Encrypt$ and $\Decrypt$, with a common key $\symmetricKey$.  $\Encrypt$ outputs a ciphertext $\ciphertext$, and decrypt will output the corresponding plaintext:

$$\Encrypt_\symmetricKey(\content) \to \ciphertext$$
$$\Decrypt_\symmetricKey(\ciphertext) \to \content$$

\paragraph{NIZKDLEQ and VerifyNIZKDLEQ}
These two algorithms are described in \autoref{proc:nizk_verif}.

$\NIZKDLEQ(q, P_1, Q_1, P_2, Q_2, \secretKey)$ creates a proof that $\log_{P_1} Q_1 = \log_{P_2} Q_2 = \secretKey$ without revealing \secretKey to the verifier.
Here $P_i$ and $Q_i$ are elements in a (prime-order) group of size~$q$ and \secretKey is an integer modulo~$q$.
This can be extended easily to a batched version by proving the same relationship for a random linear combination of points $\{P_i\}, \{Q_i\}$. For full details, see \cite{davidson2018privacy}.
\begin{figure}[ht!]
\begin{pcvstack}[center,space=1mm]
\procedureblock[colspace=-0.8cm,colsep=0ex,codesize=\footnotesize,bodylinesep=1ex]{\small$\NIZKDLEQ(q, P_1, Q_1, P_2, Q_2, \secretKey )$}{
    \pcln k \gets \scalarField \\
    \pcln A \gets k \cdot P_1, B \gets k \cdot P_2 \\
    \pcln c \gets \Hash(q || P_1|| Q_1 || P_2 || Q_2|| A|| B) \\
    \pcln s \gets k - c \cdot \secretKey \bmod q \\
    \pcln \return{\proof \gets (c, s)}
}
\vspace{1ex}
\procedureblock[colspace=-0.8cm,colsep=0ex,codesize=\footnotesize,bodylinesep=1ex]{\small$\Verify\NIZKDLEQ([q, P_1, Q_1, P_2, Q_2], z)$}{
    \pcln (c, s) \gets \proof \\
    \pcln A' \gets s \cdot P_1 + c \cdot Q_1 \\
    \pcln B' \gets s \cdot P_2 + c \cdot Q_2 \\
    \pcln \return{c \stackrel{?}{=} \Hash(q || P_1 ||  Q_1 || P_2 || Q_2|| A'|| B')}
}
\end{pcvstack}
\caption{ $\NIZKDLEQ$ and $\Verify\NIZKDLEQ$ algorithms.}
\label{proc:nizk_verif}
\end{figure}

\paragraph{Privacy Pass} In \autoref{proc:PP_issue} and \autoref{proc:PP_verify}, we define the Privacy Pass token issuance and redemption protocols from~\cite{davidson2018privacy}.
We abuse the notation for $\Hash$ in Privacy Pass to return a group element, unlike in $\NIZKDLEQ$.
For the group, we use the prime order group on Curve25519 and the standard Elligator~2 algorithm for $\Hash$~\cite{bernstein2013elligator}, as implemented in the \texttt{ed25519‑dalek} Rust crate.

\begin{figure}[ht!]
\procedureblock[colspace=-0.2cm,colsep=0ex,codesize=\scriptsize,bodylinesep=1ex]{\small\textsf{PP.BlindTokenIssue}}{
    \textsf{Client}(\publicParams, \publicKey) \<\< \textsf{Issuer}(\publicParams, \secretKey, \publicKey) \< \\ 
    \pcln (q, \generator) \gets \publicParams \<\<\< \\
    \pcln \blinding \gets_\$ \ZZ^*_q \<\<\< \\
    \pcln \nonce \gets_\$ \{0,1\}^* \\
    \pcln \blinded \gets \blinding \cdot \Hash(\nonce) \\
    \< \hspace{-1cm}\xrightarrow{\hspace{0.7cm}\mbox{\scriptsize$\blinded$}\hspace{0.7cm}} \\
    \< \< \pcln (q, \generator) \gets \publicParams \< \\
    \< \< \pcln \blindSenderToken \gets \secretKey \cdot \blinded  \< \\
    \< \< \pcln \mathsf{\pi} \gets \NIZKDLEQ(\\ \<\< \hspace{6ex} q, \generator, \publicKey, \blinded, \blindSenderToken, \secretKey) \\
    \< \hspace{-1cm}\xleftarrow{\hspace{0.7cm}\mbox{\scriptsize$(\blindSenderToken, \pi)$}\hspace{0.7cm}} \\
    \pcln \IfNotThen{\Verify\NIZKDLEQ(\\ \hspace{6ex}[q, \generator, \publicKey, \blinded, \blindSenderToken], \proof)}{\return{\bot}} \\
    \pcln \mathsf{\sigma} \gets \frac{1}{\blinding} \cdot \blindSenderToken \\
    \pcln \return{(\nonce, \sigma)}
}
\caption{Privacy Pass blind token issuance protocol.}
\label{proc:PP_issue}
\end{figure}

\begin{figure}[ht!]
\procedureblock[colspace=0.5cm,colsep=0ex,codesize=\scriptsize,bodylinesep=1ex]{\small\textsf{PP.BlindTokenVer}}{
    \textsf{Redeemer}(\nonce, \mathsf{\sigma}) \<\< \textsf{Issuer}(\secretKey) \< \\ 
    \< \hspace{0cm}\xrightarrow{\hspace{0.7cm}\mbox{\scriptsize$(\nonce, \sigma)$}\hspace{0.7cm}} \\
    \< \< \pcln \mathsf{\sigma}' \gets \secretKey \cdot \Hash(\nonce) \\
    \< \< \pcln \return{\mathsf{\sigma}' \stackrel{?}{=} \mathsf{\sigma}}
}
\caption{Privacy Pass blind token redemption protocol.}
\label{proc:PP_verify}
\end{figure}

\appendixsection{Security Games} \label{sec:security_games}

\newcommand{\gameresizeboxwidth}{0.85\linewidth}
\newcommand{\gamehalfpagewidth}{0.5\linewidth}
\newcommand{\gametextsize}{} 

\paragraph{Score Integrity}
We define score integrity in the case of a malicious $\accountabilityServer$ and an honest sender.
For simplicity, we assume no reporter privacy (no DP).
The sender starts a tag issuance protocol in oracle $\OIssueTag$.
For this oracle the adversary inputs the channel it wants the honest sender to use.
The game keeps track of the state of the sender. 
The tag issuance is complete in the $\OReport$ oracle.
$\OReport$ also includes the report protocol.
In this game, some receivers are honest while some others are corrupted. 
When a tag is issued for an honest receiver who does not report, $\accountabilityServer$ gets no information about the completed issuance, meaning that $\OReport$ is not called.
We capture the corrupted receivers with the $\OReport$ oracle, which outputs $(\opening, \proof, \blindSenderToken)$, where $\accountabilityServer$ gets $(\proof, \blindSenderToken)$ to make an honest report. 
Both corrupted receivers' reports and honest receivers' reports are modeled by $\OReport$ and count for one report.

Note that malicious senders play no role in the game. 
The adversary can run them, play with malicious \accountabilityServer and establish endorsed channels with honest or malicious receivers.
If honest receivers do not report, malicious senders play no role without using any oracles.

The goal of the adversary is to produce a false report count $\reportCount$ larger than the true number of reports, and $\reportCount$ valid pairwise different Privacy Pass tokens. 
We can prove that the one-more unforgeability (OMUF) of Privacy Pass tokens implies score integrity.
Indeed, an adversary that plays the score integrity game can be transformed into an adversary playing the OMUF game, with similar complexity and advantage.

\begin{figure*}
\gametextsize
\resizebox{\gameresizeboxwidth}{!}{
\begin{minipage}{\textwidth}
\centering
\begin{varwidth}[t]{\gamehalfpagewidth}
    \begin{algorithmic}[1]
    \item[\underline{$\mathsf{ScIntegrity}(\adv)$}]
    \State $\publicParams \gets \ppSetup(1^\lambda) $
    \State $\publicKey, \mathsf{state} \gets \adv(\publicParams)$
    \State $(\epochSecretKey_\senderMarker, \epochPublicKey_\senderMarker)\gets \ppKeyGen(\publicParams)$
    \For { $j \in \{1, \ldots, \vkLimit \}$}
    \State $(\signingKey_\senderMarker^{j} , \verificationKey_\senderMarker^{j}) \gets\sigkeygen(1^{\lambda})$
    \EndFor
    \State $0 \gets \mathsf{ctr}$ 
    \State $\nonce_1, \signature_1, \ldots, \nonce_\reportCount, \signature_\reportCount \gets \adv^{\mathsf{oracles}}(\mathsf{state}, \epochPublicKey_\senderMarker)$ \\
    \algorithmicif~ {$\reportCount \leq \mathsf{ctr}$}
    \algorithmicthen~ \textbf{abort} \\
    \algorithmicif~ {$\exists i \neq j: \nonce_i = \nonce_j$}
    \algorithmicthen~ \textbf{abort} \\
    \algorithmicif~{$\exists i: \OVerify(\nonce_i, \signature_i) \rightarrow \mathsf{false}$}
    \algorithmicthen~ \textbf{abort}
    \State \Return $1$
    \algstore{pointer}
    \end{algorithmic}
\end{varwidth}
\quad
\begin{varwidth}[t]{\gamehalfpagewidth}
    \begin{algorithmic}[1]
    \algrestore{pointer}
    \item[\underline{$\OVerify(\nonce, \signature)$}] 
    \State \Return $\ppBlindTokVer(\publicParams, \nonce, \signature, \epochSecretKey_\senderMarker)$
    \medskip
    \item[\underline{$\OIssueTag(j, \handle_\recipientMarker)$}]
    \State $(\commitment, \opening) \gets \sender.\senderReportTokenGen(\publicParams, \epochSecretKey_\senderMarker, \verificationKey_\senderMarker^{j}, \handle_\recipientMarker)$ 
    \State $\text{store state } (\verificationKey_\senderMarker^{j}, \handle_\recipientMarker, \commitment, \opening) \text{ in a fresh index } i$
    \State \Return $(\commitment, i)$
    \medskip
    \item[\underline{$\OReport(i, \token)$}] \\
    \algorithmicif~{$\text{index } i \text{ does not exist or has already been reported}$}
    \algorithmicthen~ \textbf{abort}
    \State $(\proof, \blindSenderToken) \gets \sender.\senderReportTokenGen \text{ on index } i \text{ with } \token$
    \State $\text{increment } \mathsf{ctr}$
    \State \Return $(\opening, \proof, \blindSenderToken)$
    \end{algorithmic}
\end{varwidth}
\end{minipage}}
\caption{Score integrity.}
\label{fig:repIntegrity}
\end{figure*}

We consider and adversary $\mathcal{A}$ playing the $\mathsf{ScIntegrity}$ game~\autoref{fig:repIntegrity}, with the scheme in \autoref{sec:protocol}.
We model $\Hash$, used in $\NIZKDLEQ$, with a random oracle.

The epoch secret $\epochSecretKey$ is not used in the initial phase of the tag issuance, which is in $\OIssueTag$, but is only used in the $\OReport$ and $\OVerify$ oracles.
The latter exists in the OMUF game.
The former needs to be simulated with the help of a Privacy Pass tag issuance oracle $\mathsf{Sign}(\blinded)\to \blindSenderToken$.
That oracle takes a query $\blinded$ as input and returns the response $\blindSenderToken = \epochSecretKey \cdot \blinded$.
However, $\epochSecretKey$ is used at two other places in $\senderReportTokenGen$:
to test $\ppEphemeralKey \neq \epochSecretKey \cdot \generator'$ and to compute $\mathsf{NIZKDLEQ}(q,\generator',\ppEphemeralKey,\blinded,\blindSenderToken,\epochSecretKey)$.

For the $\ppEphemeralKey \neq \epochSecretKey \cdot \generator'$ test, we have two options.
First, we can use an additional $\mathsf{OVer}(\generator',\ppEphemeralKey)$ oracle in the OMUF game, which would answer whether $\ppEphemeralKey = \epochSecretKey \cdot \generator'$ or not for a  given input $(\generator',\ppEphemeralKey)$.
Note that such an oracle can replace the $\OVerify$ oracle, which does the same, when provided with a hash preimage $\nonce$ of the first input.
Privacy Pass remains one-more unforgeable, even when the adversary is provided such an oracle $\mathsf{OVer}$.
Alternately, we can modify the $\senderReportTokenGen$ protocol and have
$\accountabilityServer$ return $s$ to the sender, so that our simulator can do the
test without using $\epochSecretKey$.
We rather adopt the first option, because it is costless.

To compute $\NIZKDLEQ$, we use the NIZK simulator, which runs the random oracle:
to compute a proof $\pi$ for parameters $(q, \generator',\ppEphemeralKey,\blinded,\blindSenderToken)$, we pick $\pi=(c,s)$ at random and we program $\Hash$ so that $\Hash(q,\generator',\ppEphemeralKey,\blinded,\blindSenderToken,A',B')=c$ for $A'=s\cdot \generator'+c\cdot \ppEphemeralKey$ and $B'=s\cdot \blinded + c\cdot \blindSenderToken$.
The simulation is perfect, except if the query to $\Hash$ was already made before, which happens with probability bounded by $q^{-1}$.

Next, line~2 and 5 of the $\mathsf{ScIntegrity}$ game and the oracles are now run by an adversary $\mathcal{B}$, who calls the $\mathsf{OVer}$ and $\mathsf{OSign}$ oracles.
We obtain an OMUF game with additional $\mathsf{OVer}$ oracle (\autoref{fig:pp_omuf}).
Clearly, the number of calls to $\mathsf{OSign}$ is equal to the $\mathsf{ctr}$ and the winning conditions are the same.
Hence, $\mathcal{A}$ playing $\mathsf{ScIntegrity}$ with simulated $\NIZKDLEQ$ and $\mathcal{B}$ playing OMUF have the same advantage.
The complexity overhead in $\mathcal{B}$ results from the simulation and is small.
Besides storing states, it consists of running a commitment, verifying a signature, running $\NIZKDLEQ$, and simulating the random oracle by
lazy sampling.
We assume that the overhead is linear in the complexity of $\mathcal{A}$.

\begin{figure*}
\gametextsize
\resizebox{\gameresizeboxwidth}{!}{
\begin{minipage}{\textwidth}
\centering
\begin{varwidth}[t]{\gamehalfpagewidth}
    \begin{algorithmic}[1]
    \item[\underline{$\mathsf{OMUF}(\mathcal{B})$}]
    \State $\publicParams \gets \ppSetup(1^\lambda) $
    \State $(\epochSecretKey_\senderMarker, \epochPublicKey_\senderMarker)\gets \ppKeyGen(\publicParams)$
    \State $\mathsf{ctr} \gets 0$ 
    \State $n_1, \signature_1, \ldots, n_\reportCount, \signature_\reportCount \gets \mathcal{B}^{\mathsf{oracles}}(\epochPublicKey_\senderMarker)$\\
    \algorithmicif~{$\reportCount \leq \mathsf{ctr}$}
    \algorithmicthen~ \textbf{abort} \\
    \algorithmicif~ {$\exists~ i \neq j, ~n_i = n_j$}
    \algorithmicthen~ \textbf{abort} \\
    \algorithmicif~{$\exists~ i~ \OVerify(n_i, \signature_i)~ \mathsf{false}$}
    \algorithmicthen~ \textbf{abort} 
    \State \Return $1$
    \algstore{pointer}
    \end{algorithmic}
\end{varwidth}
\quad
\begin{varwidth}[t]{\gamehalfpagewidth}
    \begin{algorithmic}[1]
    \algrestore{pointer}
    \item[\underline{$\mathsf{OVer}(\generator', \ppEphemeralKey)$}] 
    \State \Return $1_{\ppEphemeralKey=\epochSecretKey \cdot \generator'}$
    \medskip
    \item[\underline{$\mathsf{OSign}(\blinded)$}]
    \State $\text{increment}~ \mathsf{ctr}$
    \State $R \gets \epochSecretKey \cdot \blinded$
    \State \Return $\blindSenderToken$
    \end{algorithmic}
\end{varwidth}
\end{minipage}}
\caption{$\mathsf{OMUF}$ game in Privacy Pass.}
\label{fig:pp_omuf}
\end{figure*}

\begin{theorem}
  Given an adversary $\mathcal{A}$ with complexity $T$ playing $\mathsf{ScIntegrity}$ in the random oracle model, there exists an
  adversary $\mathcal{B}$ with complexity $\mathcal{O}(T)$ playing the Privacy Pass OMUF game, with an additional oracle $\mathsf{OVer}(\generator',\ppEphemeralKey)=1_{\ppEphemeralKey = \epochSecretKey \cdot \generator'}$ and such that
  \[
    \mathsf{Adv}_\mathcal{A}^\mathsf{ScIntegrity}
    \leq
    \mathsf{Adv}_\mathcal{B}^\mathsf{OMUF}
    +
    \frac1q\,.
  \]
\end{theorem}

\paragraph{Sender Unlinkability}
We assume an adversary who corrupted all receivers.
Senders and \accountabilityServer are honest.
Sender unlinkability is modeled by saying that a real game $\mathsf{SenderUnlink}$ can reduce to a game $\mathsf{SenderUnlink}^*$, with an ideal scheme with indistinguishable outcome.
The idea is to show that leakage about the sender is solely due to the reputation.

We define the real-world game in \autoref{fig:realWorldSenderUnlink} and the ideal-world game in \autoref{fig:idealWorldSenderUnlink}.
We denote by $\handle_\senderMarker$, the index of a new $\verificationKey_\senderMarker$ which is what the receiver sees.

\begin{figure*}
\gametextsize
\resizebox{\gameresizeboxwidth}{!}{
\begin{minipage}{\textwidth}
\centering
\begin{varwidth}[t]{\gamehalfpagewidth}
    \begin{algorithmic}[1]
    \item[\underline{$\mathsf{SenderUnlink}(\adv)$}]
    \State $J \gets 0$
    \State $((\symmetricKey,\secretKey),\publicKey, \publicParams) \gets \setup(1^\lambda)$
    \State $\adv^\mathsf{oracles}(\publicParams,\publicKey)$
    \State \Return \text{the view of } $\adv$

    \medskip
    \item[\underline{$\mathsf{OSetTime}(\timestamp)$}] \\
    \algorithmicif~ {$\timestamp<\mathsf{current\_time}$}
    \algorithmicthen~ \textbf{abort}
    \State set $\mathsf{current\_time}\gets\timestamp$
    \State \Return
    \medskip
    \item[\underline{$\mathsf{ONewSender}(\identity_\senderMarker)$}] \\
    \algorithmicif~ {$\identity_{\senderMarker}$ \text{already exists}}
    \algorithmicthen~ \textbf{abort}
    \State \text{create an account for} $\identity_\senderMarker$
    \State \Return
    \medskip
    \item[\underline{$\mathsf{OEpoch}$}] ~
    \State run the end of epoch procedure on \accountabilityServer
    \State \Return
    \medskip
    \item[\underline{$\mathsf{OSignKey}$}] ~
    \State \text{increment} $J$
    \State $(\signingKey_\senderMarker^{J}, \verificationKey_\senderMarker^{J}) \gets \sigkeygen(1^{\lambda})$
    \State \Return ~$\verificationKey_\senderMarker^{J}$
    \algstore{pointer}
    \end{algorithmic}
\end{varwidth}
\quad
\begin{varwidth}[t]{\gamehalfpagewidth}
    \begin{algorithmic}[1]
    \algrestore{pointer}
    \item[\underline{$\mathsf{ONewSenderAddr}(j_1, \ldots, j_n, D)$}] \\
    \algorithmicif~{\text{any} $\identity\{j_\ell\}$ \text{exists} }
    \algorithmicthen~\textbf{abort}
    \State $\identity_\senderMarker \gets D$ (sample a sender following distribution $D$)
    \State set $\identity\{j_\ell\} \gets \identity_\senderMarker$ \text{ for } $\ell = \{1, \ldots, n\}$
    \State \Return
    \medskip
    \item[\underline{$\mathsf{OKeyRegistration}(\identity_\senderMarker)$}] ~
    \State make $\identity_\senderMarker$ register a new $\epochPublicKey_\senderMarker$ to \accountabilityServer
    \State \Return
  \medskip
  \item[\underline{$\mathsf{OTag}(j, \handle_\recipientMarker, \timestamp)$}] \\
  \algorithmicif~ {$\timestamp \not\approx\mathsf{current\_time}$}
  \algorithmicthen~ \textbf{abort} 
  \State $\identity_\senderMarker \gets \identity\{j\}$
  \State make sender $\identity_\senderMarker$  
  \State $(\token, \opening, \proof, \blindSenderToken) \gets \senderReportTokenGen(\verificationKey_\senderMarker^{j}, \handle_\recipientMarker, \identity_\senderMarker, \timestamp)$
  \State \Return $(\token, \opening, \proof, \blindSenderToken)$
  \medskip
  \item[\underline{$\OReport(\token, \proof,\blindSenderToken)$}] 
  \State run $\accountabilityServer.\tokenTagReport(\token, \proof,\blindSenderToken)$
  \State \Return
\end{algorithmic}
\end{varwidth}
\end{minipage}}
\caption{Real-world sender unlinkability game.}
\label{fig:realWorldSenderUnlink}
\end{figure*}

\begin{figure*}
\gametextsize
\resizebox{\gameresizeboxwidth}{!}{
\begin{minipage}{\textwidth}
\centering
\begin{varwidth}[t]{\gamehalfpagewidth}
    \begin{algorithmic}[1]
    \item[\underline{$\mathsf{SenderUnlink}^*(\adv)$}]
    \State $J \gets 0$
    \State $((\symmetricKey,\secretKey),\publicKey, \publicParams) \gets \setup(1^\lambda)$
    \State $\mathsf{ctr} \gets 0$, $\mathsf{tag} \gets []$
    \State $\adv^\mathsf{oracles}(\publicParams,\publicKey)$
    \State \Return \text{the view of } $\adv$

    \medskip
    \item[\underline{$\mathsf{OSetTime}(\timestamp)$}] \\
    \algorithmicif~ {$\timestamp<\mathsf{current\_time}$}
    \algorithmicthen~ \textbf{abort}
    \State set $\mathsf{current\_time}\gets\timestamp$
    \State \Return
    \medskip
    \item[\underline{$\mathsf{ONewSender}(\identity_\senderMarker)$}] \\
    \algorithmicif~ {$\identity_{\senderMarker}$ \text{already exists}}
    \algorithmicthen~ \textbf{abort} 
    \State \text{create an account for} $\identity_\senderMarker$
    \State \Return
    \medskip
    \item[\underline{$\mathsf{OEpoch}$}] ~
    \State run the end of epoch procedure on \accountabilityServer
    \State \Return
    \medskip
    \item[\underline{$\mathsf{OSignKey}$}] ~
    \State \text{increment } $J$
    \State $(\signingKey_\senderMarker^{J}, \verificationKey_\senderMarker^{J}) \gets \sigkeygen(1^{\lambda})$
    \State \Return ~$\verificationKey_\senderMarker^{J}$
    \algstore{pointer}
    \end{algorithmic}
\end{varwidth}
\quad
\begin{varwidth}[t]{\gamehalfpagewidth}
    \begin{algorithmic}[1]
    \algrestore{pointer}
    \item[\underline{$\mathsf{ONewSenderAddr}(j_1, \ldots, j_k, D)$}] \\
    \algorithmicif~{\text{any} $\identity\{j_\ell\}$ \text{exists} }
    \algorithmicthen~\textbf{abort} \\
    \algorithmicif~{$\identity\{j_\ell\}$ \text{ exists }}
    \algorithmicthen~ \textbf{abort}
    \State $\identity_\senderMarker \gets D$ (sample a sender following distribution $D$)
    \State set $\identity\{j_\ell\} \gets \identity_\senderMarker$\text{ for } $\ell \in \{ 1, \ldots, n \}$
    \State \Return
    \medskip
    \item[\underline{$\mathsf{OKeyRegistration}(\identity_\senderMarker)$}] ~
    \State make $\identity_\senderMarker$ register a new $\epochPublicKey_\senderMarker$ to \accountabilityServer
    \State \Return
  \medskip
  \item[\underline{$\mathsf{OTag}^*(j, \handle_\recipientMarker, \timestamp)$}] \\
  \algorithmicif~ {$\timestamp \not\approx\mathsf{current\_time}$}
  \algorithmicthen~ \textbf{abort}
  \State $\identity_\senderMarker \gets \identity\{j\}$
  \State make sender $\identity_\senderMarker$  
  \State $(\token, \opening, \proof, \blindSenderToken) \gets \senderReportTokenGen(\verificationKey_\senderMarker^{j}, \handle_\recipientMarker, \identity_\senderMarker, \timestamp)$
  \State increment $\mathsf{ctr}$
  \State $\mathsf{tag}[\mathsf{ctr}] \gets (\token,\proof,\blindSenderToken)$
  \State get $\reputation_\senderMarker$ from $\token$
  \State \Return $\reputation_\senderMarker$
  \medskip
  \item[\underline{$\mathsf{OReport}^*(i)$}] \\
  \algorithmicif~ {$\mathsf{tag}[i]$ \text{does not exist}}
  \algorithmicthen~ \textbf{abort}
  \State run $\accountabilityServer.\tokenTagReport(\mathsf{tag}[i])$
  \State \Return
\end{algorithmic}
\end{varwidth}
\end{minipage}}
\caption{Ideal-world sender unlinkability game.}
\label{fig:idealWorldSenderUnlink}
\end{figure*}

The difference is that only the reputation $\reputation_\senderMarker$ leaks from sending a message.
The rest of the tag remains hidden.
Thus, the only leakage is from $\reputation_\senderMarker$.
The adversary receives no other information from the oracles.
In this ideal game, the adversary can only report genuine tokens by telling the sequence number $i$ of the creating $\OSend$ oracle call.

We start with an adversary $\adv$ playing $\mathsf{SenderUnlink}$.
First, we reduce to an adversary only submitting valid tokens and observe that, due to the EUF-CMA security of the signature scheme, all submitted tokens must have been genuinely generated.
This has two consequences.
The first one is that the adversary can index the tag with its sequence number $i$ as used by the $\OReport^*$ interface.
The second consequence is that what is encrypted in the tag is known by the game that henceforth needs no decryption.

Second, we reduce to no collision at encryption time and we replace the symmetric encryption scheme by a random permutation, which can be simulated without knowing what is encrypted.

Third, we simulate $\NIZKDLEQ$ by programming the random oracle.
Then, we can replace $\ppEphemeralKey$ from $\token$ by a random group element and skip the verification $\ppEphemeralKey \neq\epochSecretKey_\senderMarker\cdot \generator'$ by the sender.
Assuming that DDH problem is hard, the adversary does not notice the change.
We obtain a game in which the $\OSend$ oracle creates tags without using $\epochSecretKey_\senderMarker$ or $\epochPublicKey_\senderMarker$ at all.
Hence, when provided with $\reputation_\senderMarker$, the adversary can just simulate this creation as well.
We define the adversary $\mathcal{B}$ playing $\mathsf{SenderUnlink}^*$ and obtain the following result.

\begin{theorem}
  We assume that the signature is EUF-CMA secure, that the encryption is a PRP, and that the DDH problem is hard.
  For any adversary $\adv$ playing $\mathsf{SenderUnlink}$, there exists an adversary $\mathcal{B}$ playing $\mathsf{SenderUnlink}^*$ of similar complexity, and such that the outcome of both games are indistinguishable.
\end{theorem}

\ifinappendix
\appendixsection{Properties of the Script Space}\label{sec:props-of-script-space}
\else
\paragraph{Properties of the script space}
\fi
Our results in \autoref{sec:optimality} require that \messageSet is ``large'' in the following sense.
\begin{definition}[Ample script space]\label{def:ample-script-space}
\messageSet is \textit{ample} if for any script $\content\in \messageSet$ it is possible to find arbitrarily many other scripts with the same $\mathbb{E}[\probReward(\content, \cdot )\rewardLabel(\content)]$ and $\mathbb{E}[\probReport(\content, \cdot)]$.
\end{definition}
Since each script encodes the receiver's address, it suffices that for any receiver it is possible to find arbitrarily many other receivers that are expected to behave ``similarly.''
For example, a salesperson may always be able to reach arbitrarily many customers who are expected to react similarly to any specific sales script.

Even though an ample script space must be infinite, we require it to still be ``compact'' in a sense that allows a sender to decide which script to execute.
\begin{definition}[Compact script space]\label{def:compact-script-space}
\messageSet is \textit{compact} if, for any $\reputation_\senderMarker$, $\max_{\content \in \messageSet} \widetilde{\rewardLabel}(\content, \reputation_\senderMarker)$ exists.
\end{definition}
While we allow $\messageSet$ to evolve with time, we require that this change is not, in a sense, bad for the sender.
\begin{definition}[Growth assumption] \label{def:growth_assumption}
\messageSet satisfies the \textit{growth assumption} if, for any $\reputation_\senderMarker$, $\max_{\content \in \messageSet} \widetilde{\rewardLabel}(\content, \reputation_\senderMarker)$ increases (not necessarily strictly) with time.
\end{definition}
This is a natural assumption, as senders are likely to instead find more efficient scripts to replace old ones.

\appendixsection{Scaling Noise in Differential Privacy} \label{appendix:scale_noise}
We recall the differential privacy notion from \autoref{sec:anonymous_reporting}:
for any two adjacent subsets $\mathcal{D}_0, \mathcal{D}_1 \subseteq \mathcal{M}$, where $\mathcal{M}$ denotes a list of channels, and any subset $\mathcal{S}$ of possible noisy report counts,
\[ \Pr[\mathfrak{R}(\mathcal{D}_0) \in \mathcal{S}] \leq e^{\epsilon} \Pr[\mathfrak{R}(\mathcal{D}_1) \in \mathcal{S}] + \delta\,. \]

\accountabilityServer samples a noise $\noise \gets \noiseDistr$ and adds \noise to the true report count $|\mathcal{D}_0|$ (resp. $|\mathcal{D}_1|$) or $|\mathcal{D}_0| + \noise$ (resp. $|\mathcal{D}_0| + \noise$). Let $\vkLimit = \mathcal{D}_0- \mathcal{D}_1$. Let $\mathcal{S}_0 = \mathcal{S} - |\mathcal{D}_0|$.

The DP notion states that for $\vkLimit = 1$, $\noise \sim \noiseDistr\left( \mu, \sigma^2, -\frac{1}{2} \right)$, $\forall \mathcal{S}_0$, $\Pr[\noise \in \mathcal{S}_0 ] \leq e^{\epsilon} \Pr[\noise \in \mathcal{S}_0 \pm 1] + \delta\,$.

To find the noise distribution parameters for $\vkLimit$, we apply the scaling as follows. Let $\noise' = - \frac{1}{2} + \vkLimit\left( \frac{1}{2} + \noise \right)$ and define $\mathcal{S}'_0$ s.t. $\mathcal{S}_0 = \frac{1}{\vkLimit}\left( \mathcal{S}'_0 + \frac{1}{2} \right) -\frac{1}{2}$.

\begin{align*}
        \Pr [\noise' \in \mathcal{S}'_0] &= \Pr \bigg[  - \frac{1}{2} + \vkLimit \bigg( \frac{1}{2} + \noise \bigg) \\ &\qquad\qquad \in  - \frac{1}{2} + \vkLimit \bigg( \frac{1}{2} + \mathcal{S}_0 \bigg) \bigg] \\
        &= \Pr[ \noise \in \mathcal{S}_0] \\
        &\leq e^{\epsilon} \Pr[\noise \in \mathcal{S}_0 \pm 1] + \delta\, \\
        &= e^{\epsilon} \Pr\bigg[  - \frac{1}{2} + \vkLimit \bigg( \frac{1}{2} + \noise \bigg) \\ &\qquad\qquad \in  - \frac{1}{2} + \vkLimit \bigg( \frac{1}{2} + \mathcal{S}_0 \pm 1 \bigg) \bigg] + \delta\, \\
        &= e^{\epsilon}\Pr [\noise' \in \mathcal{S}'_0 \pm \vkLimit] + \delta.
\end{align*}

Thus, $\noise' \sim \noiseDistr \left( -\frac{1}{2} + \vkLimit\left(\frac{1}{2} + \mu \right), (\vkLimit \cdot \sigma)^2, -\frac{1}{2} \right)$.

\appendixsection{Further Discussion on Reporter Privacy}\label{subsec:extended_privacy_discussion}

\paragraph{Problems with report privacy}
A report privacy notion, as opposed to reporter privacy, might at first seem like a decent idea.
However, consider a scammer trying to learn whether their endorsement tags sent to various email addresses go through to human receivers.
In this case, mere report privacy would cause a problem, since the scammer can send endorsement tags repeatedly to the same receiver address over a longer period of time, eventually learning with higher confidence that the tags are likely received and reported.
However, whether any particular tag was reported is irrelevant to the scammer.
The scammer can try to get a stronger response from the receiver by sending the tags along with highly inappropriate messages that the (human) receiver is likely to react to---and report.

\paragraph{Colluding senders}
Our reporter privacy notion does not protect the receiver from colluding senders.
This seems impossible to protect against, since the receiver has no way of knowing which senders may be colluding, and the receiver's behavior may correlate across all of them (\textit{e.g.}, the receiver may be reporting every tag it receives).
The only option would be to limit the total number of reports the receiver can ever submit, but this limits the receiver too much.

\paragraph{Multiple receiver addresses}
In \autoref{sec:anonymous_reporting}, we assumed that a receiver uses only a single address for endorsed channels, but in practice the receiver may want to use multiple addresses.
In this case, the concept of a channel must be changed to capture all of the receiver's addresses, \textit{e.g.}, $(\verificationKey_\senderMarker, \{\handle_\recipientMarker\})$, where the set contains now all addresses controlled by the same receiver.
The privacy definition in \autoref{subsec:def-of-report-privacy} changes as well: $\mathcal{M}$ will instead contain a list of pairs $(\verificationKey_\senderMarker, \{\handle_\recipientMarker\})$, for which tags were requested.

To implement reporter privacy in this case, the only change that is needed is that the receiver ensures that it maintains a single dictionary \TagsDict across all of its addresses, ensuring it is properly synchronized across all of the devices it uses.
Everything else works as before.

Unfortunately, this has a subtle drawback: it may allow a sender to link multiple receiver addresses together.
Namely, a sender can establish endorsed channels with multiple addresses it wants to confirm belong to the same receiver and coerce reports (by acting inappropriately) from them all.
If the addresses belong to the same receiver, the sender's score would go down less than otherwise expected.
This is why we are not providing a receiver unlinkability guarantee.

\paragraph{Trading off privacy for reporting power}
Note that a receiver can, at will, choose to ignore the safe reporting time marker $\timestamp_\text{rep}$ and thus forgo its reporter privacy protections.
If the receiver communicates on endorsed channels with multiple addresses, it can additionally choose to give up any expectations of privacy regarding its ownership of these addresses, and report the sender as many times as it can.

\appendixsection{Proof of \autoref{th:opt_strategy_1}}\label{subsec:proof-of-th1}
In addition to the random variables $\probReward(\content, \reputation_\senderMarker)$ and $\probReport(\content, \reputation_\senderMarker)$, we also need a similar random variable $\mathsf{w}(\content, \reputation_\senderMarker)$ that outputs the probability of the sender being both rewarded and (then) reported.

If a sender decides to set up a new endorsed channel and execute a script~\content, and afterwards follow an optimal strategy, exactly one of four things can happen:
\begin{enumerate}
\item With probability $\probReport(\content, \reputation_\senderMarker) -\mathsf{w}(\content, \reputation_\senderMarker)$, it receives a report and an expected reward for the rest of the game $\optFunc_{\score, \epoch, \remainingReports}^{\reportCount +1}$.
\item With probability $1 - \probReport(\content, \reputation_\senderMarker) - \probReward(\content, \reputation_\senderMarker)+ \mathsf{w}(\content, \reputation_\senderMarker)$, it receives neither a report nor a reward (the conversation led to nothing) and an expected reward for the rest of the game $\optFunc_{\score, \epoch, \remainingReports}^{\reportCount}$.
\item With probability $\probReward(\content, \reputation_\senderMarker) - \mathsf{w} (\content, \reputation_\senderMarker)$, it receives a reward $\rewardLabel(\content)$ and an expected reward for the rest of the game $\optFunc_{\score, \epoch, \remainingReports}^{\reportCount}$.
\item With probability $\mathsf{w}(\content,\reputation_\senderMarker)$, it receives a reward $\rewardLabel(\content)$ first, but then gets reported. The expected reward for the rest of the game is $\optFunc_{\score, \epoch, \remainingReports}^{\reportCount+1}$.
\end{enumerate}
The sender's expected reward for the rest of the game after executing a script \content is thus (after simple reordering of terms)
\begin{equation}
\optFunc_{\score, \epoch, \remainingReports}^{\reportCount}(\content) :=  \probReward(\content, \reputation_\senderMarker) \rewardLabel(\content) - \probReport(\content, \reputation_\senderMarker) C_{\score, \epoch, \remainingReports}^\reportCount + \optFunc_{\score, \epoch, \remainingReports}^{\reportCount}\,,
\label{eq:local_opt_func}
\end{equation}
where $C_{\score, \epoch, \remainingReports}^\reportCount := \optFunc_{\score, \epoch, \remainingReports}^{\reportCount} - \optFunc_{\score, \epoch, \remainingReports}^{\reportCount + 1}$ measures the penalty from an additional report in the current epoch.
Hence, $\mathsf{w}(\content, \reputation_\senderMarker)$ disappears.
Furthermore, if in an optimal strategy the sender executes a script $\content$ with $(\score, \epoch, \remainingReports, \reportCount)$, then $\mathbb{E}\optFunc_{\score, \epoch, \remainingReports}^{\reportCount} (\content) = \optFunc_{\score, \epoch, \remainingReports}^{\reportCount}$ and \eqref{eq:local_opt_func} implies 
\begin{equation}
\mathbb{E}\optFunc_{\score, \epoch, \remainingReports}^{\reportCount}(\content) = \widetilde{\rewardLabel}(\content, \reputation_\senderMarker) + \optFunc_{\score, \epoch, \remainingReports}^{\reportCount+1}\,.
\label{eq:intermediary}
\end{equation}

Instead of setting up any new endorsed channels, the sender can choose to wait until the next epoch.
Therefore,
\begin{equation}
\optFunc_{\score, \epoch, \remainingReports}^{\reportCount} = \max\{\optFunc_{\reputationFunc(\score, \reportCount + \noise_\epoch), \epoch-1, \remainingReports -\reportCount - \noise_\epoch}^{0}, \max_{\content}\mathbb{E}\optFunc_{\score, \epoch, \remainingReports}^{\reportCount}(\content)  \}\,.
\label{eq:opt_func}
\end{equation}
The maximum over \messageSet in \eqref{eq:opt_func} exists because we assumed \messageSet to be compact (\autoref{def:compact-script-space}).

\paragraph{Following an optimal strategy}
We define
\begin{align}
\begin{split}
 \adversaryStrategy_{\score, \epoch, \remainingReports} := \big\{ \reportCount \mid &\reportCount + \noise_\epoch \leq \remainingReports \text{ and } \exists \content \text{ s.t. } \\ &\mathbb{E}\optFunc_{\score, \epoch, \remainingReports}^{\reportCount}(\content) \geq \optFunc_{\reputationFunc(\score, \reportCount + \noise_\epoch), \epoch-1, \remainingReports- \reportCount - \noise_\epoch}^{0} \big\}
\end{split}
\label{eq:stategy_set}
\end{align}

\textit{i.e.}, the set of values for \reportCount for which there still exists a script that the sender should execute rather than wait until the next epoch.
Clearly each $\adversaryStrategy_{\score, \epoch, \remainingReports}$ is finite (or empty).
Let $X_{\score, \epoch, \remainingReports}$ denote the smallest non-negative integer not in $\adversaryStrategy_{\score, \epoch, \remainingReports}$.
If $X_{\score, \epoch, \remainingReports} = 0$ (\textit{i.e.}, $0 \not\in \adversaryStrategy_{\score, \epoch, \remainingReports}$), the sender's optimal strategy is simply to not execute any scripts in the current epoch.
Otherwise, an optimal strategy is to execute a script $\content^*$, such that
\begin{align}
\content^* = \argmax_{\content} \mathbb{E}\optFunc_{\score, \epoch, \remainingReports}^\reportCount(\content) \stackrel{\text{Eq.}~\eqref{eq:intermediary}}{=} \argmax_\content \widetilde{\rewardLabel}(\content, \reputation_\senderMarker) \,.
\label{eq:optimal_msg}
\end{align}

Since $\content^*$ does not depend on \reportCount, for any $\reportCount \in [0, X_{\score,\epoch, \remainingReports} - 1]$,
\begin{equation}
\optFunc_{\score, \epoch, \remainingReports}^{\reportCount} \stackrel{\text{Eq.}~\eqref{eq:opt_func}}{=}  \mathbb{E}\optFunc_{\score, \epoch, \remainingReports}^{\reportCount}(\content^*) \stackrel{\text{Eq.}~\eqref{eq:intermediary}}{=} 
\widetilde{\rewardLabel}(\content^*, \reputation_\senderMarker) + \optFunc_{\score, \epoch, \remainingReports}^{\reportCount + 1}  \,.
\label{eq:recursive_reward}
\end{equation}
Applying \eqref{eq:recursive_reward} repeatedly, we get
\begin{align}\begin{split}\label{eq:optimal-strategy}
\optFunc_{\score, \epoch, \remainingReports}^{0} =  &X_{\score, \epoch, \remainingReports} \widetilde{\rewardLabel}(\content^*, \reputation_\senderMarker) \\ &+ \optFunc_{\reputationFunc(\score, X_{\score, \epoch, \remainingReports} + \noise_\epoch), \epoch-1,  \remainingReports - X_{\score, \epoch, \remainingReports} - \noise_\epoch}^{0} \,.
\end{split}\end{align}

Still, it is not clear how to concretely reach the per-epoch expected reward $X_{\score, \epoch, \remainingReports} \widetilde{\rewardLabel}(\content^*, \reputation_\senderMarker)$ in~\eqref{eq:optimal-strategy}.
To do this, the sender chooses a sequence of scripts $\overline{\content}^* = (\content_1^*,\content_2^*,\ldots)$, where each $\content_i \in \overline{\content}^*$ is optimal in the sense of \eqref{eq:optimal_msg}, and has the same $\mathbb{E}[\probReward(\content_i^*, \reputation_\senderMarker)\rewardLabel(\content_i^*)]$ and $\mathbb{E}[\probReport(\content_i^*, \reputation_\senderMarker)]$.
It is not hard to see that the expected total reward from executing the scripts in $\overline{\content}^*$ until a report occurs is given by the random variable
\begin{align*}
\mathsf{T}( \overline{\content}^*, \reputation_\senderMarker) := &\sum_{i=1}^\infty \bigg[ \Big( \sum_{j=1}^i \probReward(\content_j^*,\reputation_\senderMarker) \rewardLabel(\content_j^*)\Big)  \\ &\qquad \times \probReport(\content_i^*, \reputation_\senderMarker)\prod_{j=1}^{i-1} (1-\probReport(\content_j^*, \reputation_\senderMarker)) \bigg].
\end{align*}
Using the receivers' rationality assumption (\autoref{def:receivers-rationality}) and the fact that the $\content_i^*$ are pairwise independent, we can compute its expectation value
\begin{align*}
\mathbb{E}[\mathsf{T}( \overline{\content}^*, \reputation_\senderMarker)] := & \mathbb{E}[\probReward(\content^*,\reputation_\senderMarker) \rewardLabel(\content^*)] \mathbb{E}[\probReport(\content^*,\reputation_\senderMarker)] \\ &\quad
\times \sum_{i=1}^\infty i (1-\mathbb{E}[\probReport(\content^*, \reputation_\senderMarker))]^{i - 1} \\ = &\widetilde{\rewardLabel}(\content^*, \reputation_\senderMarker).
\end{align*}

It remains to repeat this process $X_{\score, \epoch, \remainingReports}$ times to reach the per-epoch reward in~\eqref{eq:optimal-strategy}, finally explaining \autoref{def:normalized_strategy}.

We have now proved that there exists an optimal strategy that is normalized, but in fact any optimal strategy must be normalized.
First, note that a sender executing an optimal strategy will never end an epoch after executing a script that was not reported, because such an event does not change the sender's state and hence is entirely immaterial to its strategy.
Since per report the sender's maximal expected reward is $\widetilde{\rewardLabel}(\content^*, \reputation_\senderMarker)$, it is obvious that the sender needs to reach this per each report it gets.
On the other hand, from \eqref{eq:local_opt_func} we see that if the sender's optimal strategy is to execute a script \content, then $\widetilde{\rewardLabel}(\content^*, \reputation_\senderMarker) = C_{\score, \epoch, \remainingReports}^\reportCount$.
But since the left-hand side does not depend on \reportCount, neither does the right-hand side.
This means that for each script the sender executes $\widetilde{\rewardLabel}(\content^*, \reputation_\senderMarker)$ must be the same.
Hence, an optimal strategy necessarily consist of executing scripts optimal in the sense of \eqref{eq:optimal_msg} within each epoch, until a certain number of reports are received. In other words, the strategy is normalized.
This concludes the proof of \autoref{th:opt_strategy_1}.
\appendixsection{Proof of \autoref{th:opt_strategy_2}}\label{subsec:proof-of-th2}
A rational sender always follows an optimal strategy, hence would always follow a normalized strategy.
Following a normalized strategy is easy, but the integers~$X_i$ are not bounded by anything and are hard for the sender to determine.
Our goal is to find a bounded region for the $X_i$ that is guaranteed to contain an optimal (not necessarily unique) strategy.

Let $\widehat{X}_i := \bonus - \noise_i$.
Fix any score function $\reputationFunc_\bonus^{\maxScore, \reportScale}$ (\autoref{def:score_function}) and any reputation function $\reputation$ (\autoref{def:reputation_function}).
Consider a sender's strategy where, for some $\epoch \geq 1$, $X_i \leq \widehat{X}_i$ for all $i > \epoch$ and $X_\epoch > \widehat{X}_\epoch$.
We will prove that there exists always at least as good of a strategy, where also $X_\epoch \leq \widehat{X}_\epoch$. 

Without loss of generality, let $\epoch \geq 1$ and $\mathcal{O}$ a normalized strategy $(X_\epoch, \ldots, X_0)$, where $X_\epoch > \widehat{X}_\epoch$.
This is depicted in the top branch of \autoref{fig:two_strategies}.
The lower branch denotes a different strategy $\mathcal{O}'$, where we execute $\content_\epoch^*$\footnote{By $\content_i^*$ we denote here an optimal script for the epoch~$i$, with the understanding that each executed script is actually different, as in \autoref{def:normalized_strategy}.} only until $\widehat{X}_\epoch$ reports are received and in the following epoch execute $\content_{\epoch - 1}^*$ until the same total number of reports, $X_\epoch - \widehat{X}_\epoch + X_{\epoch - 1}$, are received.
We show that $\mathcal{O}'$ is at least as good as~$\mathcal{O}$.

\newcommand{\strategyfixarrowlen}{3.2cm}
\begin{figure}
    \begin{center}
        \begin{tikzpicture}[node distance=2cm]
            \node (first) [draw, rectangle] {$\score_\epoch$};
            \node (str1_second) [draw, rectangle, above right=0.2cm and 1.1cm of first] {$\score_{\epoch - 1}$};
            \node (str1_third) [draw, rectangle, right=\strategyfixarrowlen of str1_second] {$\score_{\epoch - 2}$};
            \node (str2_second) [draw, rectangle, below right=0.2cm and 1.1cm of first] {$\score_{\epoch - 1}'$};
            \node (str2_third) [draw, rectangle, right=\strategyfixarrowlen of str2_second] {$\score_{\epoch - 2}'$};
            
            \node (dots) [below right=0.3cm and 0.8cm of str1_third] {$\cdots$};

            \draw[->] (first) -- node[sloped, anchor=center, above] {\footnotesize $(\content_\epoch^*, X_\epoch)$} (str1_second);
            \draw[->] (str1_second) -- node[above] {\footnotesize $(\content_{\epoch - 1}^*, X_{\epoch - 1})$} (str1_third);
            \draw[->] (first) -- node[sloped, anchor=center, below] {\footnotesize $(\content_\epoch^*, \widehat{X}_\epoch)$} (str2_second);
            \draw[->] (str2_second) -- node[above, align=center] {\footnotesize $(\content_{\epoch - 1}^*, X_\epoch - \widehat{X}_\epoch + X_{\epoch - 1})$} (str2_third);
            \draw[->] (str1_third) -- node[sloped] {} (dots);
            \draw[->] (str2_third) -- node[sloped] {} (dots);
        \end{tikzpicture}
    \end{center}
    \caption{The top branch depicts the strategy $\mathcal{O}$. The bottom branch depicts another strategy $\mathcal{O}'$ that differs from $\mathcal{O}$ in two epochs.
    The notation $(\content, \reportCount)$ indicates a strategy of executing $\content$ until $\reportCount$ reports are received.}
    \label{fig:two_strategies}
\end{figure}

To prove that $\mathcal{O}'$ at least as good as $\mathcal{O}$, first, we observe that $X_\epoch + \noise_\epoch > \widehat{X}_\epoch + \noise_\epoch = \bonus$.
Thus,
\begin{align*}
\score_{\epoch - 1}' &= \reputationFunc(\score_{\epoch}, \widehat{X}_\epoch + \noise_\epoch) \\
&\stackrel{\text{Eq.}~\eqref{eq:upd_reports_are_really_bad}}{\geq}
\reputationFunc(\score_{\epoch}, X_\epoch + \noise_\epoch) + \reportScale(X_\epoch -\widehat{X}_\epoch)  \\ &= \score_{\epoch - 1} + \reportScale(X_\epoch -\widehat{X}_\epoch) \,.
\end{align*}
Note that since $X_\epoch > \widehat{X}_\epoch$, $\score_{\epoch - 1}' > \score_{\epoch - 1}$.
If we denote $\reputation_i := \reputation(\score_i)$ and $\reputation_i' := \reputation(\score_i')$, then \autoref{def:reputation_function} implies $\reputation_{\epoch - 1}' \geq \reputation_{\epoch - 1}$.
Furthermore,
\begin{align*} \score_{\epoch - 2}' &\geq \reputationFunc(\score_{\epoch - 1} + \reportScale(X_{\epoch} -\widehat{X}_\epoch), X_\epoch - \widehat{X}_\epoch + X_{\epoch - 1} + \noise_{\epoch - 1}) \\ &\stackrel{\text{Eq.}~\eqref{eq:upd_reputation_vs_report}}{\geq}  \reputationFunc(\score_{\epoch - 1}, X_{\epoch - 1} + \noise_{\epoch - 1}) = \score_{\epoch - 2} \,. 
\end{align*}
This means that when the two strategies converge $\mathcal{O}'$ already yields at least as high of a score for the sender as~$\mathcal{O}$ and therefore at least as high total reward as~$\mathcal{O}$.

It remains to consider the expected reward difference in epochs $\epoch$ and~$\epoch - 1$.
A straightforward calculation shows that in these epochs difference in the expected rewards is
\begin{equation*} 
\begin{gathered}
    (X_\epoch - \widehat{X}_\epoch)\left[ \widetilde{\rewardLabel}(\content_{\epoch - 1}^*, \reputation_{\epoch - 1}') - \widetilde{\rewardLabel}(\content_{\epoch}^*, \reputation_{\epoch}) \right] \\
    + X_{\epoch - 1} \left[ \widetilde{\rewardLabel}(\content_{\epoch - 1}^*, \reputation_{\epoch - 1}') - \widetilde{\rewardLabel}(\content_{\epoch-1}^*, \reputation_{\epoch-1}) \right] 
\end{gathered}
\end{equation*}
We note that it follow immediately from the receivers' rationality assumption (\autoref{def:receivers-rationality}) that $\widetilde{\rewardLabel}(\content, \cdot)$ is an increasing function (not necessarily strictly increasing) for any script~$\content$.
Since $\reputation_{\epoch - 1}' \geq \reputation_{\epoch - 1}$, the second summand is non-negative.

The first summand is trickier to reason about.
By hypothesis, $X_\epoch > \widehat{X}_\epoch$. 
Equation \eqref{eq:upd_grace_score} implies $\score_{\epoch - 1}' \geq \score_\epoch$, so $\reputation_{\epoch - 1}' \geq \reputation_\epoch$ and by receivers' rationality
$\widetilde{\rewardLabel}(\content_{\epoch - 1}^*, \reputation_{\epoch - 1}') \geq \widetilde{\rewardLabel}(\content_{\epoch - 1}^*, \reputation_{\epoch})$,
so the difference in the first summand is greater than or equal to $\widetilde{\rewardLabel}(\content_{\epoch - 1}^*, \reputation_{\epoch}) - \widetilde{\rewardLabel}(\content_{\epoch}^*, \reputation_{\epoch})$, which is non-negative due to the growth assumption (\autoref{def:growth_assumption}).
Thus, the first summand is also non-negative and we can conclude that $\mathcal{O}'$ is at least as good as $\mathcal{O}$.
If $\mathcal{O}$ was optimal, $\mathcal{O}'$ is also optimal.

The analysis of the second-to-last epoch is a simplified version of the above. The first claim of \autoref{th:opt_strategy_2} follows by applying this result repeatedly and assuming the growth assumption holds for the entire duration of the game.

The second claim of \autoref{th:opt_strategy_2} follows from applying the first claim to the optimal strategy that exists according to \autoref{th:opt_strategy_1}.
\appendixsection{Colluding Sender and AS}\label{subsec:colluding_as_and_sender}

A natural security guarantee against a colluding malicious sender and \accountabilityServer would be a ``report transparency'' guarantee for reporting receivers.
Roughly, report transparency would ensure to reporting receivers that their reports are taken correctly into account when updating the sender's score.
First, we note that this proof would be meaningless unless each receiver verifies it against the same view of \accountabilityServer's database.
The database cannot be public, so instead \accountabilityServer would need to publish cryptographic commitments to the contents of its database, which requires a public bulletin board functionality. This problem is related to verifiable e-voting, where voters need a proof that their ballots were counted in a tally.

Another problem is that for the receiver to verify the proof, it needs to know for which sender (really, $\identity_\senderMarker$) its report should be counted towards.
However, this is in conflict with the unlinkability guarantee: the receiver cannot know~$\identity_\senderMarker$.
It would be relatively easy for the receiver to verify that its report was counted towards a score corresponding to the $\verificationKey_\senderMarker$ that the sender used with the receiver, but this is insufficient since the sender does not need to use $\verificationKey_\senderMarker$ ever again (especially, when it colludes with~\accountabilityServer).

A third problem is that negative noise cannot be used for reporter privacy, because reporting receivers must be able to find their reports taken into account.
Instead, we would need a way for \accountabilityServer to receive an appropriately distributed number of valid sender-tokens that it can use as (positive) noise.
\appendixsection{Further Results on Optimality} \label{subsec:optimality-further-results}
\paragraph{Reward discrepancy analysis}
If the sender follows a bounded strategy $\mathcal{O} := (X_\epoch, \ldots, X_0)$, as in \autoref{th:opt_strategy_2}, how far is its expected total reward from what an optimal strategy $\mathcal{O}^\text{opt} := (X_\epoch^\text{opt}, \ldots, X_0^{\text{opt}})$ would yield?
We make the simplifying assumptions that the sender uses its entire budget for~$\remainingReports$ and starts from the maximum score~$\maxScore$ (\autoref{def:score_function}).

Since $X_i \leq \widehat{X}_i$ and $X_i^\text{opt} \leq \widehat{X}_i$, for $i\geq 1$, the sender's score never decreases from~$\maxScore$.
We denote $\reputation_\text{max} := \reputation(\maxScore)$.
In this case, a strategy $\mathcal{O}^\text{opt} = (-\noise_\epoch, \ldots, -\noise_1, -\noise_0 + \remainingReports)$ is optimal and the difference in the expected total rewards is
\begin{equation*}
\begin{gathered}
\sum_{i\geq 1} (X_i + \noise_i) \left[ \widetilde{\rewardLabel}(\content_0^*, \reputation_\text{max}) - \widetilde{\rewardLabel}(\content_i^*, \reputation_\text{max}) \right] \\
\leq \epoch \bonus \left[ \widetilde{\rewardLabel}(\content_0^*, \reputation_\text{max}) - \widetilde{\rewardLabel}(\content_\epoch^*, \reputation_\text{max}) \right].
\end{gathered}
\end{equation*}
In the general case, the sender's ability to minimize this difference depends strongly on the stability of \messageSet over time, on the sender's willingness to postpone its gain with the hope that more rewarding scripts will be available later in the game, and on the tolerance level~\bonus.

In the special case where \messageSet remains the same throughout the game, each $\content_i^* = \content_0^*$ and the difference becomes zero: we have an optimal strategy.
This means that it does not matter when exactly the optimal scripts $\content_0^*$ are executed, as long as per each epoch the sender's report count is between $-\noise_i$ and $\widehat{X}_i = \bonus - \noise_i$.
Note that a trivial strategy, where the sender waits until the very last epoch to act, is not optimal even in this case.

\paragraph{Sender's game with delayed reporting}
Our analysis in \autoref{subsec:proof-of-th1} and \autoref{subsec:proof-of-th2} assumes endorsement tags have an unrealistic expiration time parameter $\expTime = 0$, where tag expiration times do not span even one full epoch.
Recall from \autoref{subsec:delayed_reporting} that this means \accountabilityServer receives the report immediately in the same epoch where the tag was sent and subsequently uses the report as input to the score function at the end of the epoch.

However, the analysis for $\expTime \geq 1$ is not that different.
After making the perturbation as in the strategy $\mathcal{O}'$ in \autoref{fig:two_strategies}, we need to proceed as in $\mathcal{O}$ for \expTime additional epochs to make a claim analogous to $\Delta_\text{rest} \geq 0$.
This is because it is not enough anymore to just show that (analogously to \autoref{fig:two_strategies}), $\score_{\epoch - 2 - \expTime}' \geq \score_{\epoch - 2 - \expTime}$; we also need to show that the next update uses the same number of reports in both strategies.
With this minor change, the proof strategy we used for \autoref{th:opt_strategy_2} can be repeated almost verbatim and the theorem statement applies without change.

\appendixsection{Related Work: Reputation Systems}\label{subsec:reputation_systems}
There is a vast amount of literature on reputation systems, including \cite{josang2002beta,farmer2010building,steinbrecher2006design,kerschbaum2009verifiable,goodrich2011privacy,clauss2013k,schiffner2011limits,blomer2015anonymous,schaub2016trustless,zhai2016anonrep,bemmann2018fully,blomer2018practical,bag2018privacy,hauser2023street}.
Good surveys can be found in~\cite{josang2007survey,hendrikx2015reputation}.
Following the definitions of~\cite{josang2002beta}, reputation systems have two main components: an ``engine'', determining reputation values, and a ``propagation mechanism'', detailing how reputation scores are accessed.
Some reputation engines are monotone, which means reputation changes only in one direction (typically up), whereas others allow both increase and decrease of reputation.
Propagation mechanisms can be either centralized or decentralized.
Some propagation mechanisms update reputation in real time, while others wait for sufficient votes for privacy reasons~\cite{kerschbaum2009verifiable}.
Bidirectional reputation systems has raters and ratees that are both allowed to rate each other.
It is common that the ratings are kept in an escrow until both parties have left their ratings.
Unidirectional reputation systems only support rating in one direction.

In this terminology, \oursystem is a centralized unidirectional reputation system.
Its engine is described by a score function that satisfies \autoref{def:score_function}; a concrete example is given in \autoref{eq:our_upd}.
The engine is ``somewhat monotone'', in that our reports result only in a lowered score, but we have an in-built and fixed score recovery mechanism to counter accidental or malicious reports, and eventually recover the reputation for senders that have improved their behavior.
Our score range (\autoref{def:score_function}) is bounded from above for this reason: new users start with a fixed maximal score and try to keep it high by not engaging in conversations that are likely to get reported as inappropriate.

\paragraph{Privacy-preserving reputation systems}
Many centralized reputation systems, such as product ratings for online stores, may only provide privacy by limiting the users' view. 
Others use techniques like cryptography and secure hardware to achieve stronger guarantees, \textit{e.g.}, against a malicious service provider~\cite{steinbrecher2006design,kerschbaum2009verifiable,goodrich2011privacy,clauss2013k,blomer2015anonymous,schaub2016trustless,zhai2016anonrep,bemmann2018fully,blomer2018practical,bag2018privacy,hauser2023street}.
Comprehensive surveys of are found in \cite{gurtler2021sok,hasan2022privacy}.
Hasan \textit{et al.}~\cite{hasan2022privacy} catalogues properties these systems tend to provide, including also various security, integrity, and transparency guarantees.

Goodrich and Kerschbaum~\cite{goodrich2011privacy} analyze incentives in bidirectional reputation systems using a game theoretic approach, but only consider strategies for a single bidirectional rating event. In contrast, we study the game throughout the sender's ``lifetime'' in \autoref{sec:optimality}.
Unlike us, they do not have a formal privacy guarantee.

Schiffner \textit{et al.}~\cite{schiffner2011limits} and Clau{\ss} \textit{et al.}~\cite{clauss2013k} present formal privacy definitions for reputation systems in the bidirectional setting.
They are also concerned about the privacy of the value of the rating, whereas every rating has the same value in \oursystem.
Their analysis is interesting, but the scenario differs too much from ours for a meaningful comparison.

Hauser \emph{et al.}~\cite{hauser2023street} explore the problem of aggregating reputation scores from multiple sources in a secure and private manner.
Their system satisfies several security and privacy properties: anonymity of users, account unlinkability, integrity and unforgeability of reputation scores, and score freshness.
Recall (\autoref{subsec:security_privact_obj}) that \oursystem provides score integrity, communication privacy, report privacy, and unlinkability.
While these are related, they do not match perfectly with the guarantees in \cite{hauser2023street} due to the significant differences in the scenarios.

Many reputation systems face coercion issues, with ratees potentially pressuring raters for favorable ratings.
For example, Resnick and Zeckhauser~\cite{resnick2002trust} found in eBay transactions that negative ratings often result in retaliatory negative ratings, undermining the reputation system's utility.
Prior solutions tend to not provide formal guarantees~\cite{petrlic2014privacy}, or operate in settings incompatible with \oursystem~\cite{clauss2013k}.
Instead, we define a new formal and practical notion of report privacy based on a differential privacy.

In a recent paper, Bader \textit{et al.}~\cite{bader2023reputation} discuss the challenges of using reputation systems for business-to-business transactions and outline a proposal based on fully homomorphic encryption.
This scenario is closely related to ours, but they present almost no details of the system.

\paragraph{Threats for reputation systems}
Reputation systems are vulnerable to various threats despite (and sometimes because of) any privacy properties they may provide~\cite{hoffman2009survey,koutrouli2012taxonomy,sekar2023technical,he2022market,movschovitz2013analysis,resnick2002trust}.
Moreover, our hypothesis is that privacy technologies alone cannot address the threats these systems face today, as reputation systems themselves are enabling and even incentivizing undesired behavior.

This is why with \oursystem we take a more comprehensive approach.
We focus on a narrow well-defined scenario where we can clearly express what the purpose of the system is, utilize modern privacy technologies, and use a sophisticated score function that we can argue provably drives participants towards the desired kind of behavior.

Denial-of-Service (DoS) attacks against reputation systems~\cite{hoffman2009survey} are difficult to protect against, and we leave them outside our threat model.
For example, \accountabilityServer could simply refuse to respond to messages from senders or from receivers.
\appendixsection{Related Work: Spam Filters}\label{subsec:spam_filters}
Spam, or indiscriminately sent unsolicited communication, has been problematic since the mid-1990s, affecting not just email but all communication systems that allow sending messages to anyone without prior authorization~\cite{krebs2014spam,cormack2008email,SchlegelSpam,grier2010spam}.
In addition to legal measures, various technical anti-spam solutions have been designed and deployed over the years, including blocklists~\cite{ramachandran2007filtering,grier2010spam}, IP reputation systems~\cite{esquivel2010effectiveness} and other more sophisticated reputation-based approaches~\cite{chirita2005mailrank,prakash2005fighting,zheleva2008trusting,zhang2009ipgrouprep}, and statistics and ML-based detection methods~\cite{guzella2009review,dada2019machine}.
Authentication and anti-spoofing systems like SPF~\cite{IETF-SPF}, DKIM~\cite{IETF-DKIM}, and DMARC~\cite{IETF-DMARC} also combat unwanted communication, although a recent study~\cite{wang2022large} found numerous problems with current adoption of particularly DKIM.

Spam filters operate on a per-message basis, whereas the aim of \oursystem is not to block anything, but provide an additional signal of trust to an entire communication channel.
While spam filters offer only limited agency to their users, \oursystem empowers senders by allowing them to append an explicit endorsement to their messages, and empowers receivers to hold senders accountable through the reporting system.
Senders have an incentive to do this, as an empowered receiver that sees a (decent) reputation score is more likely to take the sender's message seriously, whereas such incentives are missing from spam filters.
In \oursystem senders can view their scores, which is not the case with spam filters.
The receiver's view is similar to ``report spam'' buttons, but the reporter's impact is more direct, described explicitly by a public score function.
Another important difference is that many anti-spam techniques rely on a PKI or long-term identity keys, further complicating their adoption especially for small companies, whereas \oursystem does not in principle require a PKI.

\paragraph{Proof-of-work and money burning}
Proof-of-work has repeatedly been suggested as an effective countermeasure for spam.
The ideas is that the sender must pay a computational cost for each message, thereby making much of mass emailing unprofitable~\cite{pricing_via_combatting_junk_mail,hashcash,pow_spam_filter}. A series of studies have arrived at different conclusions regarding whether it can (see \cite{liu2006proof}) or cannot (see \cite{pow_cant_work}) provide an effective solution.
The problem is that many legitimate senders need to send a lot of messages; blindly penalizing this is counterproductive.
Camp and Liu~\cite{pow_cant_can_does_work} argue that by combining proof-of-work with a reputation system may yield a more successful strategy.
An economic analysis of money-burning for generic mechanism design (\textit{e.g.}, for spam-fighting) was done in~\cite{money_burning_optimal_mechanism}.

All of these studies rely on an assumption that malicious actors tend to have a much higher volume of messages and lower value-per-message than honest users. 
However, introducing such cost can be unnecessarily burdensome to honest senders and therefore seems to be a poor fit for our scenario.
One of the key considerations in using \oursystem for this application was to only penalize bad behavior, \textit{i.e.}, messages getting reported.

\section{Generalized Score Function}\label{app:generalized_score_func}
In \eqref{eq:our_upd}, we presented an example of a fairly simple score function that satisfies \autoref{def:score_function}.
In addition to its simplicity, it has the convenient property that recovery per epoch is faster (specifically, $\reportCount - \bonus$), when the score is less than a threshold (zero, in this example), and slower (up to~$b \leq 1$) when the score is larger.
This disincentivizes bad behavior, where a sender with a high score behaves poorly, then waits for a short time for the score to recover, and repeats its bad behavior.
If the score recovers slower when it is higher, then the sender would have to wait an impractically long time to perform this attack, or simply settle for a lower score.
Note that this attack is not in conflict with the results in \autoref{subsec:optimality}, as this simply not the sender's optimal strategy; nevertheless, it makes sense to think of ways to reduce the expected reward from it.

Unfortunately, \eqref{eq:our_upd} is not very practical in many cases, because the speed of score recovery is always very small compared to the speed of score loss.
Namely, if the sender gives out roughly $\scoreScale$ endorsement tags per epoch, then its maximal score loss in a single epoch is $O(\scoreScale)$, whereas recovering from this requires $O(\scoreScale)$ epochs.
Next, we present a generalization of \eqref{eq:our_upd}, which allows for much more flexible recovery behavior.

\paragraph{Example score function family}
Let $\ell \in \mathbb{Z}_{\geq 1} \cup \{ \infty \}$.
Let $\recoveryExp_1\ldots\recoveryExp_\ell \in \mathbb{R}_{\geq 0}$ and $\recoveryRate_1\ldots\recoveryRate_\ell \in (0, 1]$ denote sequences of real numbers.
Let $\score_1\ldots\score_{\ell}$ denote a decreasing sequence of real numbers.
The idea is that $\recoveryRate_i$ represents the score recovery rate when $\score \in [\score_{i+1}, \score_i)$, and that these intervals are not smaller than $\reportScale$, which represents the amount of score loss from a single report (recall \autoref{def:score_function}).
Thus, to cover the entire domain, we need~$\score_1 = \maxScore$.
If $\ell < \infty$, we denote $\score_{\ell + 1} := -\infty$.
If $\ell = \infty$, we require $\cup_{i} [\score_{i+1}, \score_i) = (-\infty, M)$.
Our generalized example score function family is presented in~\eqref{eq:generalized_upd}.
\begin{figure*}
\begin{equation}
  \widetilde{\reputationFunc}(\score, \reportScale) := 
    \begin{cases}
      \score - \reportScale(\reportCount - \bonus), & \reportCount \geq \bonus \\
      \score, & \reportCount < \bonus,\linebreak\score = \maxScore \\
      \min \left\{ \score + \recoveryRate_i\left[\reportScale(\bonus - \reportCount)\right]^{\recoveryExp_i}, \score_i \right \}, & \reportCount < \bonus,\, \score \in [\score_{i+1}, \score_i)
    \end{cases}
    \label{eq:generalized_upd}
\end{equation}
\end{figure*}

To address to concern discussed at the beginning of this section, we additionally prefer that high scores recover slower than low scores.
Looking at \eqref{eq:generalized_upd}, we see that this is the case if
\begin{equation}\label{eq:extra_condition} \recoveryRate_i\left[\reportScale(\bonus - \reportCount)\right]^{\recoveryExp_i} < \recoveryRate_{i+1}\left[\reportScale(\bonus - \reportCount)\right]^{\recoveryExp_{i+1}} \end{equation}
for any $i$ and $\reportCount < \bonus$.

Note that \eqref{eq:our_upd} is a special case of \eqref{eq:generalized_upd}, where $\reportScale = 1$, $\ell = 2$, $\recoveryExp_1 = 0$, $\recoveryExp_2 = 1$, $\recoveryRate_1 = \recoveryRate$, $\recoveryRate_2 = 1$, $\score_1 = \maxScore$, $\score_2 = 0$, and $\score_3 = -\infty$.

In \autoref{th:score_function_th} we show that, under some limitations on the parameters, \eqref{eq:generalized_upd} satisfies \autoref{def:score_function}.
We also show that, under further limitations, it satisfies the condition in~\eqref{eq:extra_condition}.

\begin{theorem}\label{th:score_function_th}
\autoref{eq:generalized_upd} satisfies \autoref{def:score_function} if:
\begin{enumerate}
\item $\score_{i+1} < \score_i - \reportScale$, for each $i=1\ldots\ell$.
\item $0 \leq \recoveryExp_i \leq 1$ and $0 < \recoveryRate_i \leq \reportScale^{1-\recoveryExp_i}$, for each $i=1\ldots\ell$.
\end{enumerate}
Furthermore, \eqref{eq:generalized_upd} satisfies \eqref{eq:extra_condition} if, for each $i = 1 \ldots \ell$, $\recoveryExp_{i} \leq \recoveryExp_{i + 1}$ and $\recoveryRate_i \reportScale^{\recoveryExp_i} < \recoveryRate_{i+1} \reportScale^{\recoveryExp_{i+1}}$.
\end{theorem}

It is straightforward to check that \eqref{eq:generalized_upd} satisfies properties \eqref{eq:upd_reports_are_really_bad} and~\eqref{eq:upd_grace_score}.

It is also straightforward to check \eqref{eq:upd_report_are_bad} for each branch of \eqref{eq:generalized_upd} independently.
This comes down to the fact that $(\bonus - \reportCount)^{\recoveryExp_i} \geq (\bonus - \reportCount - 1)^{\recoveryExp_i}$, when $\reportCount < \bonus$, and keeping in mind that $\bonus$ and $\reportCount$ are integers.

Checking \eqref{eq:upd_reputation_vs_report} is the trickest part.
If $\reportCount \geq \bonus + 1$, \eqref{eq:upd_reputation_vs_report} obviously holds.
If $\reportCount = \bonus$, the LHS of  \eqref{eq:upd_reputation_vs_report} is simply~$\score + \reportScale$ and the RHS becomes $\min\{ \score + \recoveryRate_i\reportScale^{\recoveryExp_i}, \score_i \}$,
if $\score \in [\score_{i+1}, \score_i)$.
\autoref{eq:upd_reputation_vs_report} holds, since $0 < \recoveryRate_i \leq \reportScale^{1-\recoveryExp_i}$.
If $\reportCount < \bonus$ and $\score = \maxScore - \reportScale$, then obviously \eqref{eq:upd_reputation_vs_report} is satisfied (the equation is not applicable when $\score > \maxScore - \reportScale$).

If $\reportCount < \bonus$ and $\score \in [\score_{i+1}, \score_i - \reportScale)$ for some~$i$, then LHS and RHS of~\eqref{eq:upd_reputation_vs_report} become $\min\{ \score + \reportScale + \recoveryRate_i \left[ \reportScale (\bonus - \reportCount) \right]^{\recoveryExp_i}, \score_i \}$ and $\min\{ \score +  \recoveryRate_i \left[ \reportScale ( \bonus - \reportCount + 1 ) \right]^{\recoveryExp_i}, \score_i \}$, respectively.
Denoting $z := \bonus - \reportCount$, we then see that \eqref{eq:upd_reputation_vs_report} requires
\[ \frac{\reportScale^{1 - \recoveryExp_i}}{\recoveryRate_i} \geq (z + 1)^{\recoveryExp_i} - z^{\recoveryExp_i}. \]
If $\recoveryExp_i = 0$, this is trivially true.
If $\recoveryExp_i = 1$, this is equivalent to $\reportScale^{1 - \recoveryExp_i} \geq 
\recoveryRate_i$.
Otherwise, when $0 < \recoveryExp_i < 1$, $z\mapsto z^{\recoveryExp_i}$ is concave, so
\[ \recoveryExp_i z^{\recoveryExp_i - 1} \geq (z + 1)^{\recoveryExp_i} - z^{\recoveryExp_i}, \]
and it suffices to require
\[ \frac{\reportScale^{1 - \recoveryExp_i}}{\recoveryRate_i} \geq \recoveryExp_i z^{\recoveryExp_i - 1}. \]
But since $z\geq 1$, this holds as long as
$\reportScale^{1-\recoveryExp_i} \geq \recoveryRate_i \recoveryExp_i$,
which follows from our assumptions $\reportScale^{1 - \recoveryExp_i} \geq 
\recoveryRate_i$ and $0 < \recoveryExp_i < 1$.

If $\reportCount < \bonus$ and $\score \in [\score_i - \reportScale, \score_i)$ for some $i\geq 2$ (note that \eqref{eq:upd_reputation_vs_report} is not applicable for $i=1$), then LHS and RHS of~\eqref{eq:upd_reputation_vs_report} become $\min\{ \score + \reportScale + \recoveryRate_{i - 1} \left[\reportScale ( \bonus - \reportCount) \right]^{\recoveryExp_{i - 1}}, \score_{i - 1} \}$ and $\min\{ \score + \recoveryRate_i \left[ \reportScale (\bonus - \reportCount + 1 ) \right]^{\recoveryExp_i}, \score_i \}$, respectively.
Since obviously both $\score + \reportScale + \recoveryRate_{i - 1} \left[\reportScale ( \bonus - \reportCount) \right]^{\recoveryExp_{i - 1}} > \score_i$ and $\score_{i-1} > \score_i$,~\eqref{eq:upd_reputation_vs_report} is satisfied.

It is straightforward to verify that \eqref{eq:extra_condition} holds as long as, for each $i$, $\recoveryExp_{i} \leq \recoveryExp_{i + 1}$ and $\recoveryRate_i \reportScale^{\recoveryExp_i} < \recoveryRate_{i + 1} \reportScale^{\recoveryExp_{i + 1}}$, concluding the proof of \autoref{th:score_function_th}.

\end{document}